\documentclass[%
 reprint,
 amsmath,amssymb,
 aps,
]{revtex4-2}

\usepackage[dvipsnames]{xcolor}
\usepackage{graphicx}
\usepackage{dcolumn}
\usepackage{bm}
\usepackage{amsmath,amssymb,amsfonts}
\usepackage[caption=false]{subfig}
\usepackage{xcolor}
\usepackage{siunitx}
\usepackage{tabularx}
\usepackage{float}

\begin{document}

\preprint{APS/123-QED}

\title{Systematic Design of Transmission-type Polarization Converters Comprising Multi-layered Anisotropic Metasurfaces}
\thanks{DOI: https://doi.org/10.1103/PhysRevApplied.14.034049. Copyright © 2020 American Physical Society. All rights reserved.}

\author{Filippo Costa}
\affiliation{%
 Dipartimento di Ingegneria dell'Informazione, Università di Pisa, Pisa, 56122, Italy
}%
 \email{filippo.costa@unipi.it}
\author{Michele Borgese}%
\affiliation{%
 Alten Italia, Milano, 20134, Italy
\\}

\date{\today}

\begin{abstract}
A simple but efficient approach for the synthesis of transmission-type wideband polarization converters is presented. The proposed configuration comprises multilayer metasurfaces including resonant particles which are progressively rotated layer by layer. The progressive rotation of the particles allows for a polarization conversion over a large frequency band. The polarizing structure is efficiently designed and optimized through a transmission line model approach handling the cascade of anisotropic impedance layers and dielectrics. An optimized 8-layers design based on gradually rotated dipole resonators is presented as a proof of concept. The results obtained through the efficient transmission line model are compared with full-wave simulations once that the structure was optimized showing satisfactory agreement. A prototype of the wideband polarization converter has been fabricated and measured. 
\end{abstract}

\maketitle


\section*{Introduction}

The ability to manipulate the polarization state of electromagnetic waves is of vital importance in a wide range of applications spanning from microwave to optics. Common applications at microwave frequencies are related to communications antennas or microwave devices such as circulator and isolators. Several optical devices are also based on polarizing surfaces. Some examples are optical sensing, photography and devices relying on light manipulation. The control of the polarization of the light can be accomplished with both reflecting or transmitting polarizers. Reflection only polarization converters \cite{doumanis2012anisotropic,borgese2018optimal}, are simpler to design with respect to transmission type ones since the amplitude control is guaranteed by the presence of a ground plane which provides total reflection. In transmission type polarization converters \cite{iwanaga2008ultracompact, weis2009strongly,pfeiffer2014bianisotropic}, the
simultaneous control of both amplitude and phase is required. In practical applications, broadband performance and angular stability over a wide range are required \cite{mueller2017metasurface}.
Conventional approaches for the manipulation of the state of light at optical frequencies rely on quarter or half wave-plates \cite{hale1988stability, kruk2016invited}, which are made of birefringent materials composed of crystalline solids and liquid crystals. However, the inherent disadvantages in terms of size, collimation, and bandwidth \cite{zhao2013tailoring, samoylov2004achromatic} of these configurations prevent their miniaturization and integration of optical system. 
In the microwave region, a popular structure employed for converting linear polarized waves into circular polarized ones is based on the so called Pierrot unit cell \cite{pierrot1966elements, roy1996reciprocal}. The Pierrot unit cell is composed of two orthogonal monopoles connected by a vertical quarter-wavelength segment. Depending on the orientation of monopoles, the resonant element can act as a left-hand circular-polarization (LHCP) or right-hand circular-polarization (RHCP) selective surface. An improved version of the Pierrot cell employs closely-spaced helices \cite{morin1995circular, wang2017broadband, yang2010ultrabroadband}. However, the evident drawback of these devices is that they are three-dimensional structures and require advanced fabrication techniques which forbid their implementation in integrated systems.

A more attractive solution for designing transmission polarization converters is the use of multilayer metasurfaces without three-dimensional features \cite{glybovski2016metasurfaces}. Several examples of polarization converters based on Frequency Selective Surfaces (FSS) and metasurfaces have been proposed in the literature  \cite{costametamaterials,pfeiffer2014bianisotropic,iwanaga2008ultracompact, weis2009strongly, zhao2011manipulating,  zhu2013design, winkler2010polarization, grady2013terahertz,Eleftheriades2013_PRA,Figotin2001_PRA, Fan2018_OE,Cong2014_LPR,Cong2013_APL, Fan2015_AM, fan2018broadband, abadi2016wideband, momeni2016broadband, pfeiffer2014bianisotropic}. Some of the available configurations are designed at a single frequency \cite{pfeiffer2014bianisotropic,iwanaga2008ultracompact, weis2009strongly, zhao2011manipulating,  zhu2013design, winkler2010polarization, grady2013terahertz,Eleftheriades2013_PRA,Figotin2001_PRA, Fan2018_OE}. On the other hand, other configurations available in the literature are instead capable of converting the polarization over a broad frequency band \cite{Cong2014_LPR,Cong2013_APL, Fan2015_AM, fan2018broadband, abadi2016wideband, momeni2016broadband, pfeiffer2014bianisotropic}. Often, unconvetional shapes are employed relying on the experience of the designer \cite{Cong2013_APL,abadi2016wideband, momeni2016broadband,pfeiffer2014bianisotropic}. A systematic design procedure for the synthesis of the multilayer configuration of the polarization converter is not available in the literature.

This work presents a general design framework of  transmission-type polarization converters. The polarization converter comprises multilayer metasurfaces with an anisotropic element gradually rotated layer by layer. The optimization of the structure is based on an analytical Transmission Line (TL) model to compute the transmission and reflection response of multilayer metasurface comprising anisotropic elements. The optimized design is obtained by controlling the number of layers, the rotation factor and the thickness of each layer, as well as all the electrical parameters involved in the design.

\section{Polarization converter configurations} 
Let us consider an electromagnetic (EM) plane wave propagating along the $z$-axis in a Cartesian coordinate system and a polarizing surface located on the orthogonal $xy$-plane. In general, the EM field laying on $xy$-plane can be expressed as: 

\begin{equation} \label{eq:E_inc}
{\underline{E}}^{inc}  = {E}_x {\underline{i}}_x  + {E}_y {\underline{i}}_y 
\end{equation}

Considering ${\underline{\underline{\tau}}}$ the tensor representing the transmission properties of the material:
 \begin{equation} \label{eq:Tau}
                    \underline{\underline{\tau}}=
 \begin{bmatrix}
                    \tau_{xx} & \tau_{xy} \\[5pt]
                    \tau_{yx} & \tau_{yy}  \\
                  				\end{bmatrix}     
\end{equation}

the EM field transmitted by the polarizing surface can be expressed as follows: 

\begin{equation} \label{eq:E_t2}
\resizebox{.9\hsize}{!}{ ${\underline{E}}^{t}  = \underline{\underline{\tau}} \, {\underline{E}}^{inc}=({E}_x {\tau}_{xx}  + {E}_y {\tau}_{xy} ) {\underline{i}}_x  + ({E}_y {\tau}_{yy}  + {E}_x {\tau}_{yx} ) {\underline{i}}_y$}
\end{equation}

To convert the polarization of the impinging EM field into the orthogonal one, two different configurations are possible: symmetric and asymmetric polarization converters. The two configurations are discussed in the following subsections.

\subsection{Symmetric polarization converters} 

Let us consider a plane wave propagating toward $z$ direction and illuminating the polarization converter with an electric field ${E}_0$ and azimuthal angle $\varphi = 45^\circ$. In such a way ${E}_x$ and ${E}_y$ component of the impinging wave have the same amplitude: ${\underline{E}}^{inc}  = {E}_0 \, cos(45^\circ) \, {\underline{i}}_x  + {E}_0 \, sin(45^\circ) \, {\underline{i}}_y$ . In this case the polarization conversion can be obtained by imposing the the following conditions in the polariser tensor:
\begin{equation} \label{eq_conditions_symmetric_pol1}
\begin{cases} |\tau_{xx}|=|\tau_{yy}|=1 \\ |\tau_{xy}|=|\tau_{yx}|=0 \\ \angle(\tau_{xx}-\tau_{yy})=180^\circ \end{cases}
\end{equation}

According to (\ref{eq_conditions_symmetric_pol1}), in order to design a polarization converter based on this mechanism, the amplitudes of the $x$ and $y$ components of the transmission coefficient have to be close to unity in order to avoid losses and the phases should exhibit an offset of $180^\circ$ for the frequency range where the polarization rotation has to be accomplished.

If the conditions in (\ref{eq_conditions_symmetric_pol1}) are verified, the transmission ${\underline{\tau}}$ matrix will have the following form: 
 \begin{equation} \label{eq:T}
                    \underline{\underline{\tau}}=
 \begin{bmatrix}
                    1 & 0 \\[5pt]
                    0 & -1  \\

                  				\end{bmatrix}     
\end{equation}

Thus, the transmitted field will be ideally converted into cross-polar component:
\begin{equation} \label{eq:E_t1}
{\underline{E}}^{t}  = {E}_x {\underline{i}}_x  - {E}_y {\underline{i}}_y  
\end{equation}

The polarization rotation can be also demonstrated by  operating a coordinate rotation of $\varphi = 45^\circ$. In this new coordinate system $(x',y')$ the impinging electric field is entirely polarized along $x'$ (${\underline{E}}^{inc}  = {E}_0 \, {\underline{i}}_x'$) and the transmitted field can be computed by multiplying the impinging field by the transmission matrix represented on this new coordinate system. The new transmission matrix ${\underline{\underline{\tau}}}^\prime$ is computed by applying a rotation transformation 
to the original transmission matrix $\underline{\underline{\tau}}$ defined in the $xy$-plane:

\begin{equation}
{\underline{\underline{\tau}}}^\prime={\underline{\underline{R}}(\varphi=45^\circ)}^{-1}{\underline{\underline{\tau}}}\,{\underline{\underline{R}}(\varphi=45^\circ)}= 
                  			    \begin{bmatrix}
                    0 & 1 \\[5pt]
                    1 & 0  \\
                  				\end{bmatrix}    
\end{equation}

where: 

\begin{equation}
\underline{\underline{R}}(\varphi)=\begin{bmatrix} cos(\varphi) &-sin(\varphi) \\ sin(\varphi) & cos(\varphi) 	\end{bmatrix}
\end{equation}

In the new coordinate system, the only non-zero elements are the off-diagonal ones that are both equal to $1$. This means that the polarization converting structure is $symmetric$ (both the fields polarized along $x'$ and $y'$ are entirely converted into the opposite polarization. The transmission matrix for a backward propagation can be easily obtained after a rotation of 180$^\circ$, with respect to the $x$-axis, of the original matrix $\underline{\underline{\tau}}$.

\subsection{Asymmetric polarization converter} 

Let us illuminate the polarization converting structure, located in the $xy$-plane, with a field ${E}_0$ polarized along $x$ direction, ${\underline{E}}^{inc}  = {E}_0 {\underline{i}}_x$, (or along $y$ direction). The polarization conversion can be obtained by imposing the the following conditions in the polariser tensor:

\begin{equation} \label{eq_conditions_symmetric_pol}
\begin{cases} |\tau_{xx}|=|\tau_{yy}|=0 \\ |\tau_{yx}|=1 \;\;|| \;\;|\tau_{xy}|=1 \\ |\tau_{yx}|=0 \;\;|| \;\;|\tau_{xy}|=0 \end{cases}
\end{equation}

According to (\ref{eq_conditions_symmetric_pol}), the amplitudes of the co-polar components of the transmission coefficient must be close to zero and the amplitude of one of the two cross-polarized transmission coefficients close to $1$. No conditions about the phases have to be imposed. The maximization of $|\tau_{yx}|$ as well as the minimization of $|\tau_{xy}|$  can be obtained with chiral structures \cite{menzel2010advanced}. This polarization converter topology can be classified as \textit{asymmetric linear polarizer} \cite{Pfeiffer_PRA, zhang2013interference, menzel2010asymmetric, mutlu2012diodelike, huang2012asymmetric}. Indeed, the structure exhibits selective transmission and rotation for a specific linear polarization and the undesired polarization is completely reflected. If the structure is analysed from the opposite side, it exhibits polarization conversion properties for the opposite polarization preserving the passivity condition.

\section{Design approach} 
 A general and simple design approach for designing transmission-type polarization converters is not available in the literature. To this aim, a fast simulation tool is needed to avoid a not efficient procedure based on a full-wave electromagnetic solver. 

In order to synthesize a wideband polarization converter, the conditions provided in the previous section must be met for several frequencies simultaneously. In the symmetric case, both amplitude and phase constraints must be satisfied. In this case, the polarization converter structure does not exhibit chiral properties \cite{abadi2016wideband}. The asymmetric design is instead based on rotations of metasurfaces among different layers \cite{Pfeiffer_PRA}.  

The design tool presented on this paper is applied to asymmetric polarization converter. The metasurface layers, comprising anisotropic unit elements with arbitrary shape, are modelled through an equivalent circuit representation. The element is partially and gradually rotated layer by layer with an angle $\varphi$. 

\begin{figure}[h]
    \centering
        \includegraphics[width=0.9\linewidth]{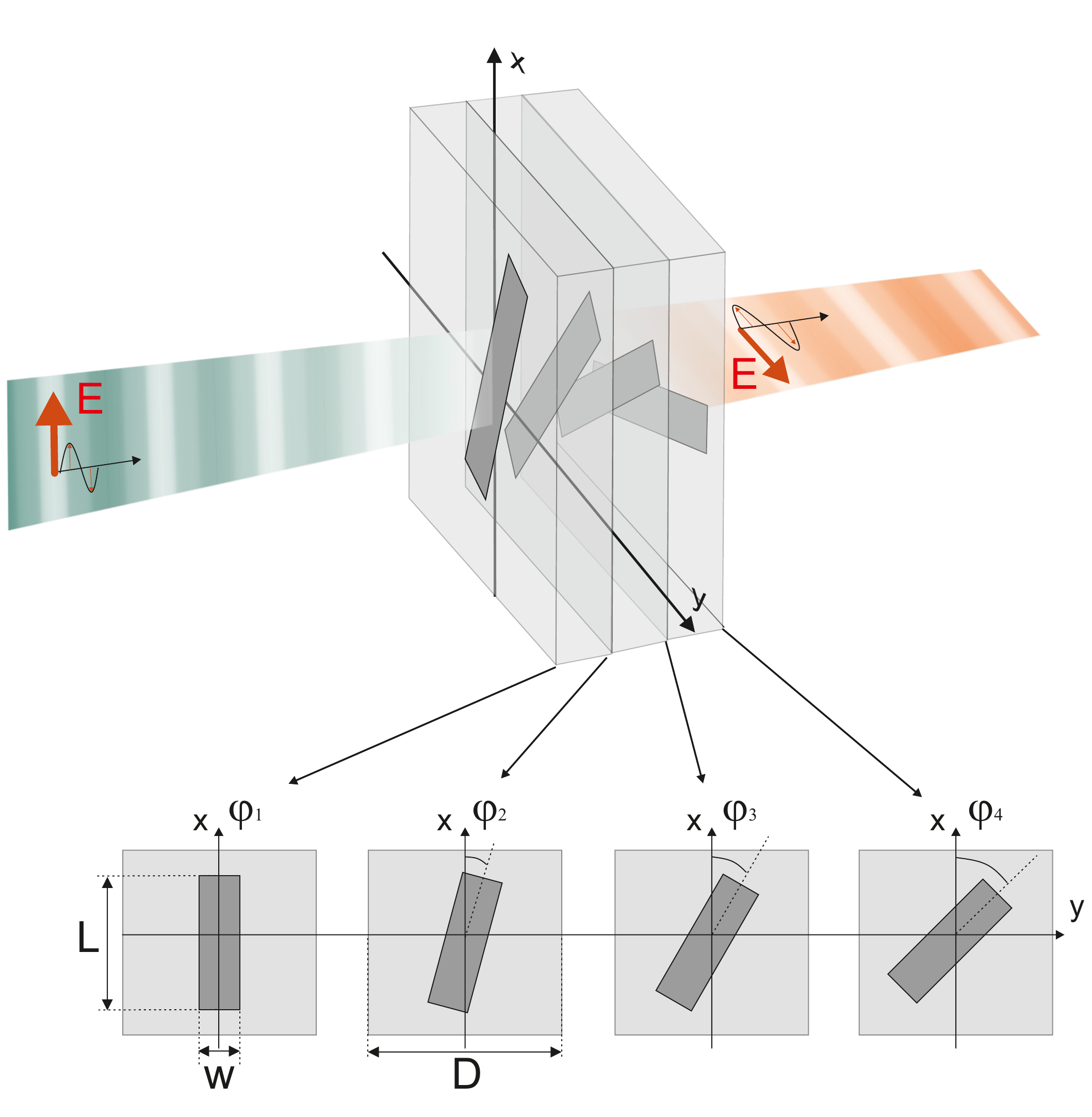}
      \caption{Layout of the polarization converter}
  \label{fig_layout} 
\end{figure}

\begin{figure}[h]
    \centering
        \includegraphics[width=0.8\linewidth]{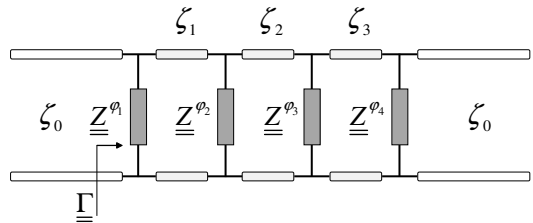}
      \caption{Equivalent circuit of the polarization converter}
  \label{fig_circuit_model} 
\end{figure}

\begin{figure} 
\centering
\subfloat[]{%
       \includegraphics[width=0.86\linewidth]{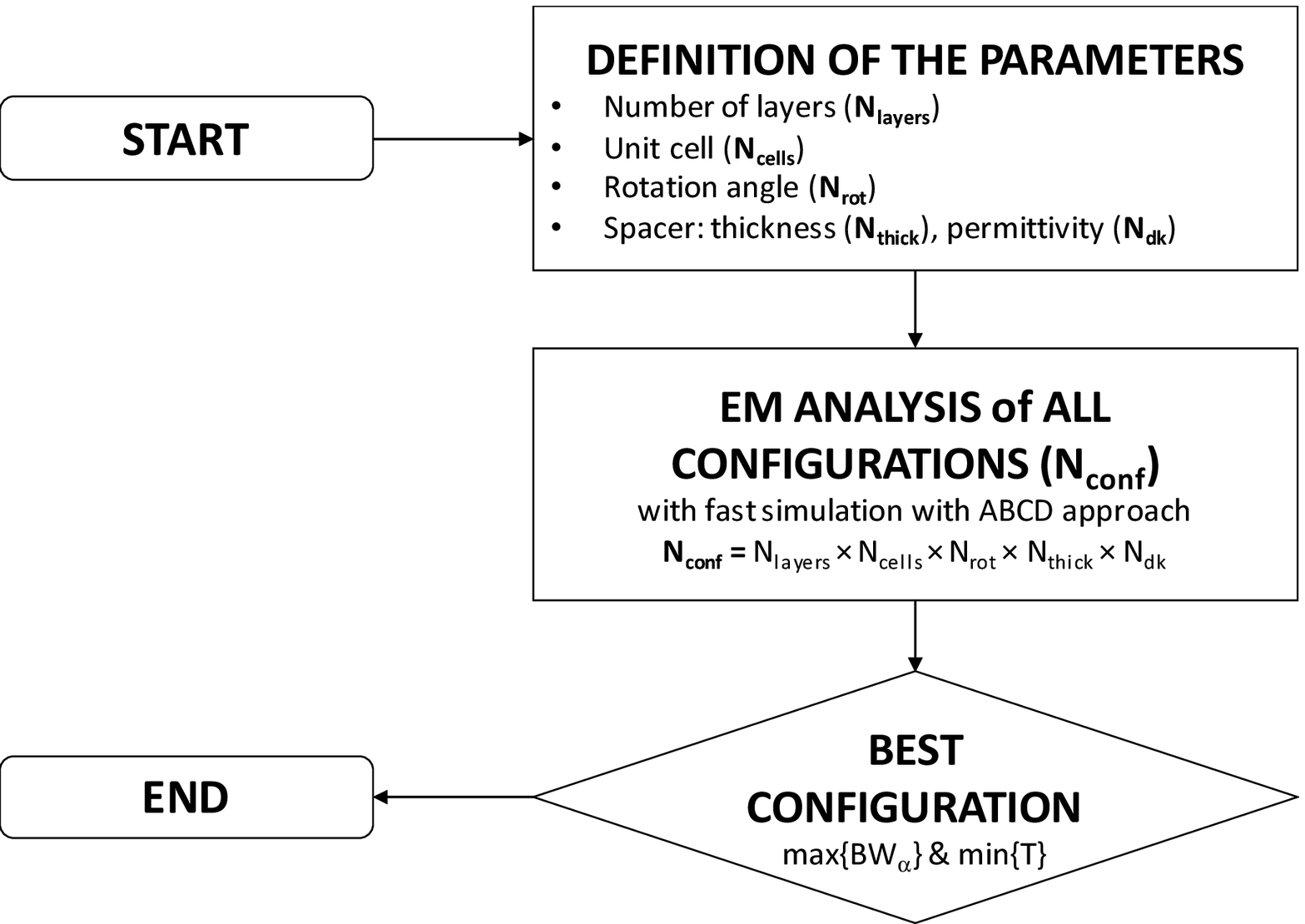}}
 \quad
 \\
  \subfloat[]{%
       \includegraphics[width=0.8\linewidth]{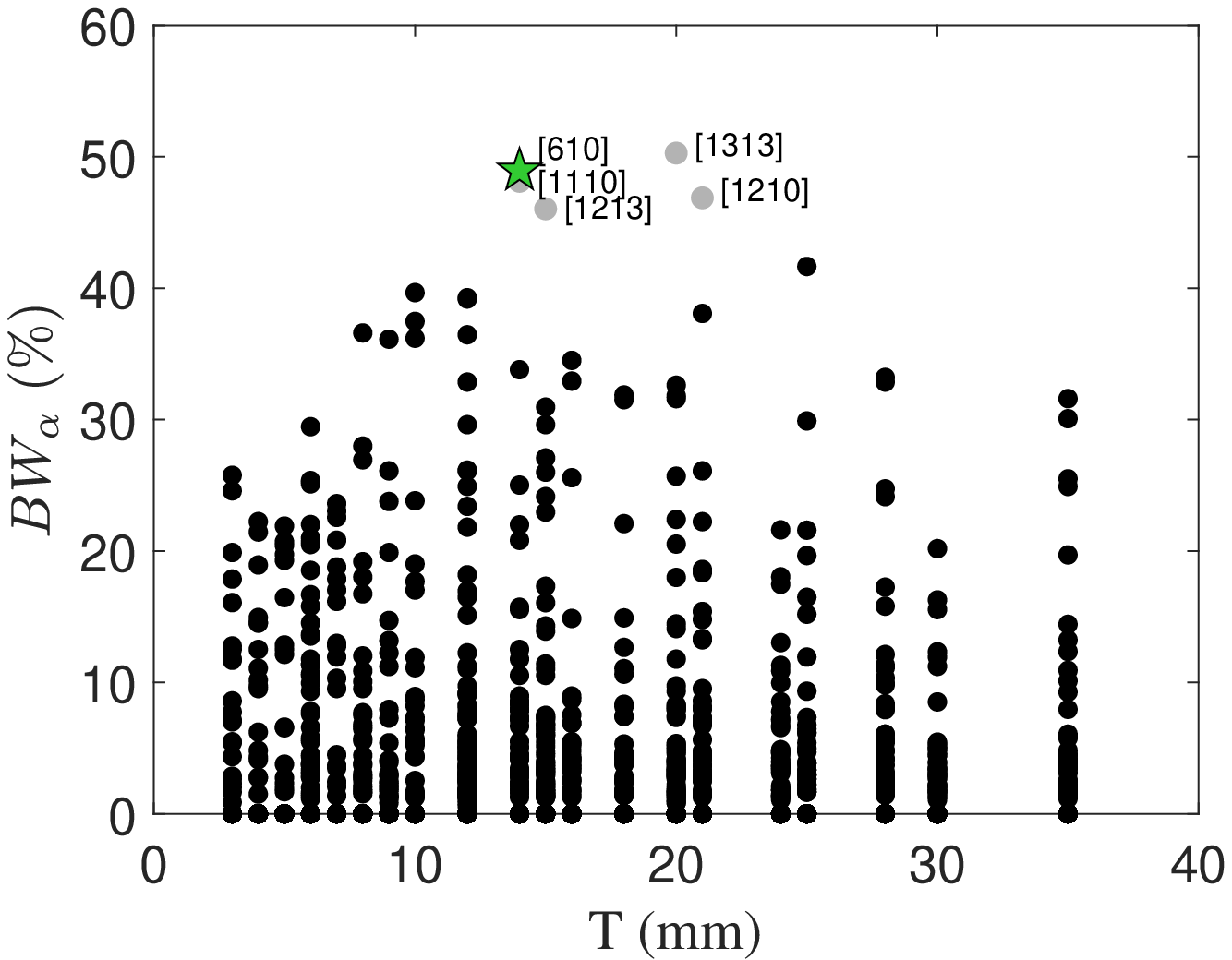}}
 \quad
 \\
  \subfloat[]{%
        \includegraphics[width=0.8\linewidth]{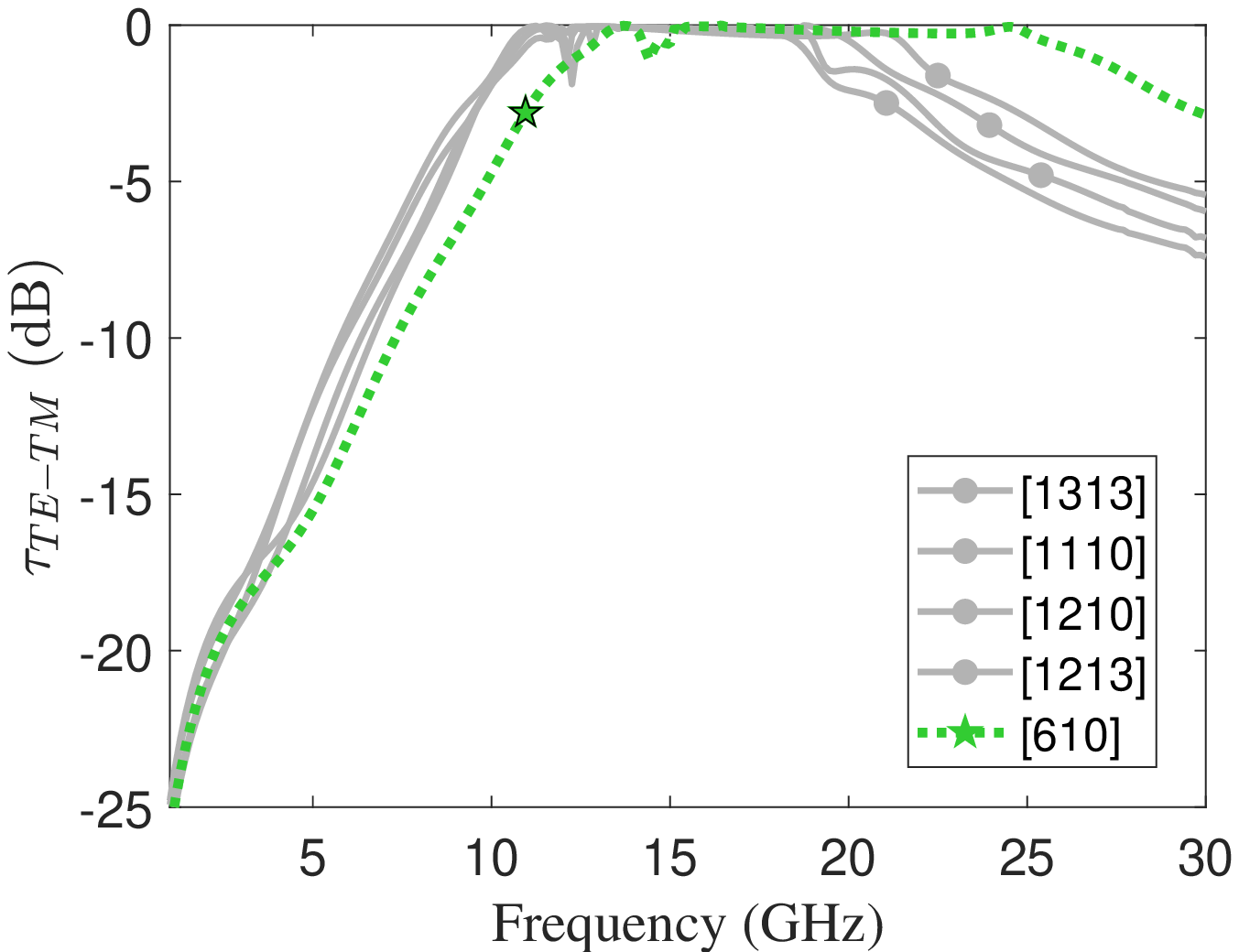}}
       \caption{(a) Block diagram of the analysis procedure; (b) analysed solutions; (c) transmission crosspolar coefficient of the best five solutions.}
  \label{fig_pareto} 
\end{figure}

The adopted design strategy is general and the analysis can be performed via an efficient transmission line (TL) model approach. The TL model, differently from a full-wave simulation based on electromagnetic solvers, allows for a very fast optimization of the multilayer structure. This approach provides accurate results if the distance between consecutive layers is large enough to avoid the interaction of the high order Floquet modes with nearby periodic surfaces. The condition is satisfied if the distance between the layers is larger than one third of the periodicity of the periodic layers \cite{costa_efficient_2012}.  Fig.~\ref{fig_layout} reports a three-dimensional layout of the multilayer structure and the partial rotation method for generic anisotropic particles. The equivalent TL model used to solve the EM problem is reported in Fig.~\ref{fig_circuit_model}. The block diagram of the optimization procedure performed to synthesize the wideband polarization converter is shown in Fig.~\ref{fig_pareto}(a). 

The analysis is performed by testing all the possible configurations according to the selected variation of the parameters in a certain interval decided by the designer. 
In order to perform the analysis, five parameters are selected: number of layers, unit cell topology, rotation angle, spacer thickness, spacer permittivity. Each parameter can assume a certain number of values within a range chosen by the designer. Subsequently, all the possible solutions are: $N_{conf}= N_{layers} \times N_{cells} \times N_{rot} \times N_{thick} \times N_{dk}$ where $N_{layers}$, $N_{cells}$, $N_{rot}$, $N_{thick}$, $N_{dk}$ are the number of values in each range of the selected parameters. During the analysis, a fitness function is computed for all the solutions after the fast analysis based on ABCD approach. The fitness function is represented by the percentage bandwidth of the crosspolar transmission coefficient $\tau_{TE-TM}$ above a certain threshold $\alpha$ ($BW_{\alpha}$). Once that the analysis of the solutions set has been carried out, the goodness is evaluated with a plot highlighting both the benefit ($BW_{\alpha}$) and the cost (total thickness $T$ of the structure) for each solution as shown in Fig.~\ref{fig_pareto}(b). The variation range of each variable is reported in Table ~\ref{tab_optimization_parameters}. The number of configurations analyzed in this example is $N_{conf} =2000$. The computation time for the analysis of the 2000 solutions is roughly 15 minutes. The five best solutions are highlighted in grey color and reported in Fig.~\ref{fig_pareto}(c) and the selected design has been marked with a star in both figures. The parameters of the five best solutions are reported in Table ~\ref{tab_best_solutions}. The selected configuration, which is identified with a star in Table ~\ref{tab_best_solutions}, is the best compromise between the total thickness and the percentage bandwidth among the analyzed ones. In the analysis process $\alpha$ has been set to $-0.4$ dB. The proposed procedure is general and the number of analyzed configurations could be also increased by preserving a reasonable computation time.

\begin{figure}
\centering
\begin{tabular}{ccccc}
\includegraphics[width=1.6cm,height=1.6cm,angle =0]{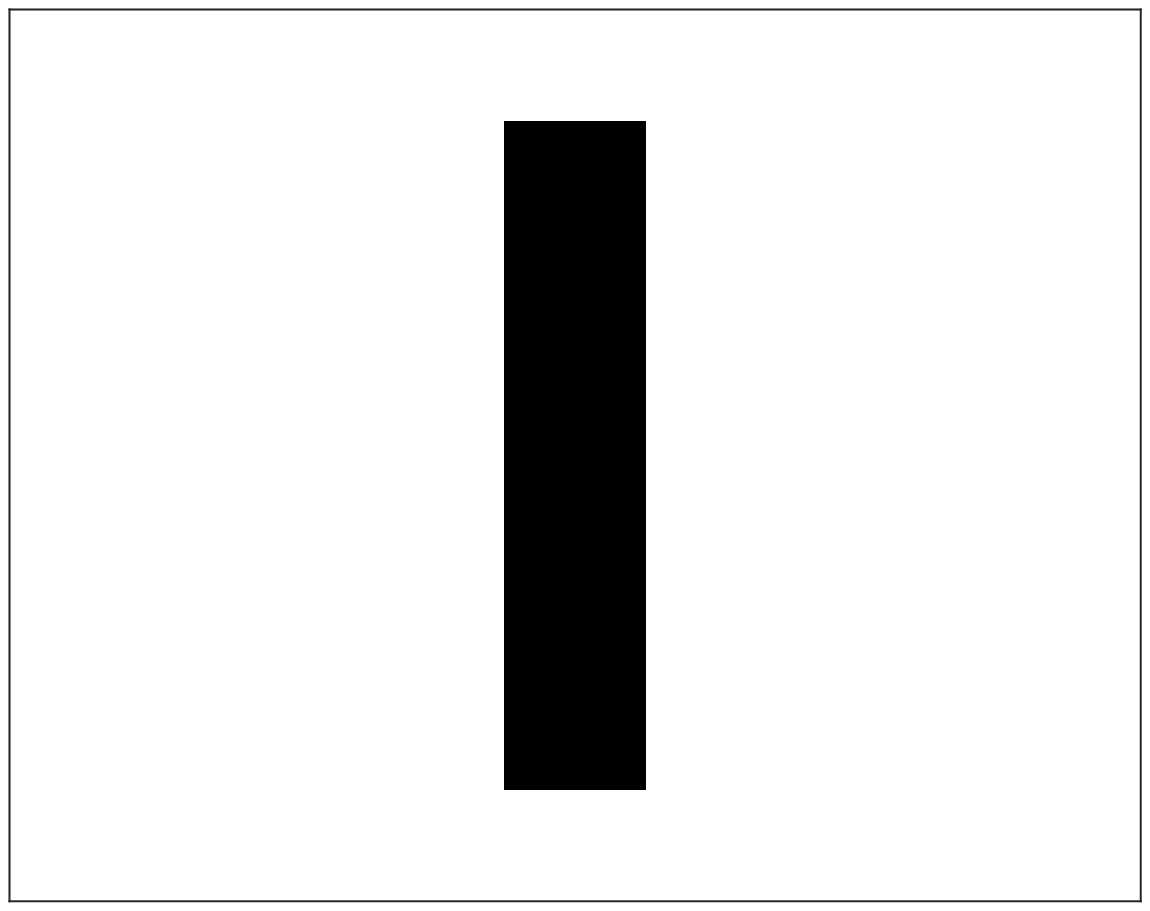} & \hspace{2mm}
\includegraphics[width=1.6cm,height=1.6cm,angle =0]{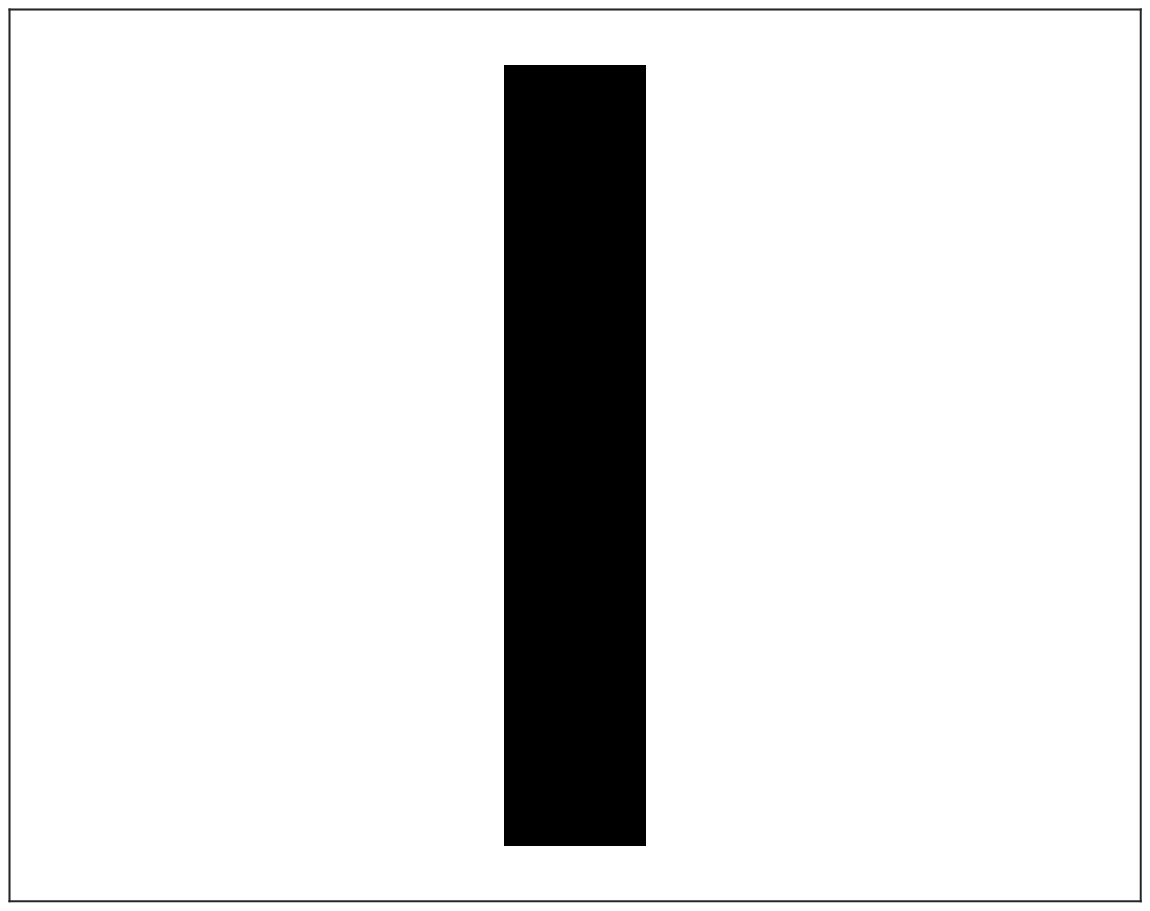} &\hspace{2mm}
\includegraphics[width=1.6cm,height=1.6cm,angle =0]{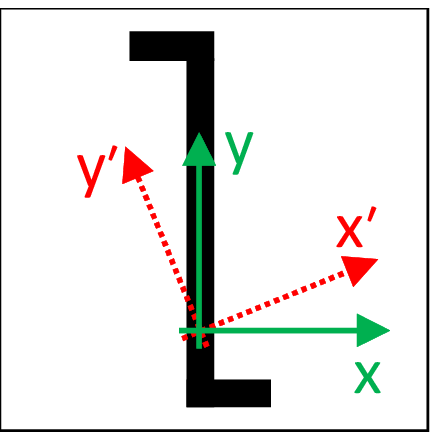} & \hspace{2mm}
\includegraphics[width=1.6cm,height=1.6cm,angle =0]{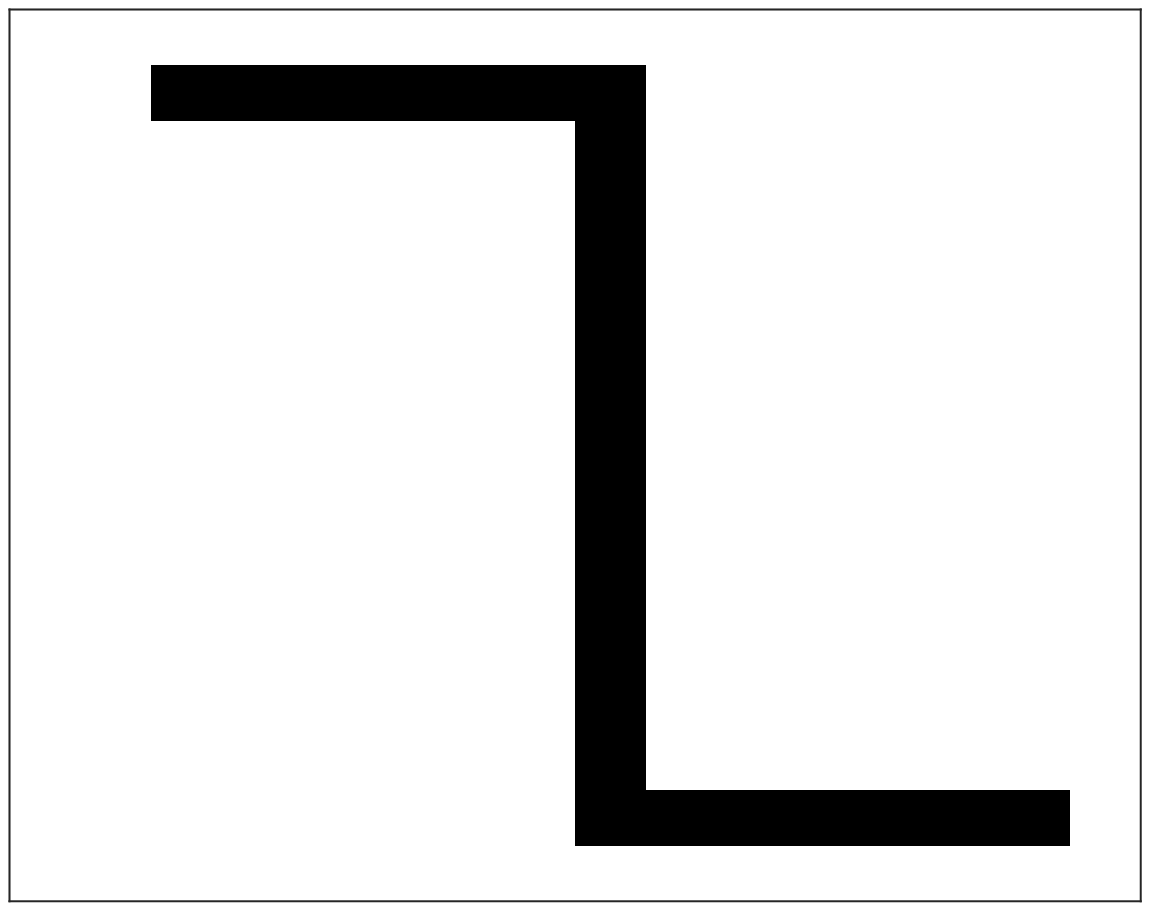} & \\
(a)  & (b) & (c) & (d) \\[6pt]
\end{tabular}
\caption{Analyzed unit cells: (a) short dipole, (b) dipole, (c) short-loaded dipole, (d) loaded dipole. In (c), the principal $(x,y)$ and rotated $(x^\prime,y^\prime)$ Cartesian reference systems are shown.}
  \label{fig_UnitCells} 
\end{figure}

\begin{table}[h]
\caption{Parameters used for the analysis reported in Fig.~\ref{fig_pareto}(b).}
\centering
\begin{tabular}{|c|c|c|}
\hline
\textbf{Parameter} & \textbf{Range} & \textbf{Number of values} \\
\hline
 spacer thickness & $[1,5]$ mm & $N_{thick} = 5$ \\
\hline
$\varepsilon_r$ & $[1,5]$ & $N_{dk} = 5$ \\
\hline
rotation angle & $[9^\circ,18^\circ]$ & $N_{rot} = 4$ \\
\hline
number of layers & $[4,8]$ & $N_{layers} = 5$ \\
\hline
unit cell topology & Fig.~\ref{fig_UnitCells}(a)-(d) & $N_{cells} = 4$ \\
\hline
\end{tabular}
\label{tab_optimization_parameters}
\end{table}

\begin{table}[H]
\caption{Parameters of the best five solutions highlighted in grey in Fig.~\ref{fig_pareto}(b).}
\centering
\begin{tabular}{|c|c|c|c|c|c|c|}
\hline
\textbf{Sol. idx} & \textbf{Cell} & $\boldsymbol{\varepsilon_r}$ & $\boldsymbol{\varphi (^\circ)}$ & \textbf{Layers} & $\boldsymbol{BW_\alpha (\%)}$ & \textbf{T (mm)} \\
\hline
[1313] & Fig.~\ref{fig_UnitCells}(c)  &  1  & 15  & 6 & 50.28  & 20\\
\hline
[610]$^\star$ &  Fig.~\ref{fig_UnitCells}(b)  &  1  & 12  & 8 & 49.94  & 14\\
\hline
[1110] &  Fig.~\ref{fig_UnitCells}(c)  &  1  & 12  & 8 & 48.12  & 14\\
\hline
[1210] &  Fig.~\ref{fig_UnitCells}(c)  &  1  & 12  & 8 & 46.87  & 15\\
\hline
[1213] &  Fig.~\ref{fig_UnitCells}(c)  &  1  & 15  & 6 & 46.03  & 15\\
\hline
\end{tabular}
\label{tab_best_solutions}
\end{table}

\section{Analysis Method} 

As previously pointed out, the analysis of the cascaded anisotropic metasurfaces relies on a transmission line model in which the metasurface is represented through a shunt impedance \cite{asadchy2015broadband}. The impedance matrix of the metasurface element as a function of frequency can be derived after a full-wave simulation for a specific azimuth angle. The metasurface impedance relates the tangential components of the electric and magnetic fields according to the following expression:

\begin{equation}
\begin{bmatrix}
                    E_{x}  \\[5pt]
                    E_{y} \\
                  				\end{bmatrix}   = \begin{bmatrix}
                    Z_{xx} & Z_{xy} \\[5pt]
                    Z_{yx} & Z_{yy}  \\

                  				\end{bmatrix}    \begin{bmatrix}
                    -H_{y}  \\[5pt]
                    H_{x} \\
                  				\end{bmatrix}  
\end{equation}

Although there are cases in which $ Z_{xy} $ and $ Z_{yx} $ are equal to zero, in general, for anisotropic metasurfaces $ Z_{xy}= Z_{yx} \neq 0 $. In this case, two separates equivalent circuit representations of the anisotropic metasurface can be computed on the \textit{x} and \textit{y} independently \cite{abadi2016wideband}. In the general case, it is possible to rotate the metasurface element on the crystal axes $(\chi_1,\chi_2)$ where these terms are equal to zero \cite{patel_TMTT_2013,selvanayagam_2014,borgesecosta2020}. Therefore, each metasurface can be characterized on the crystal axes with five parameters: $L_{\chi_1}$, $C_{\chi_1}$, $L_{\chi_2}$, $C_{\chi_2}, \varphi^\chi) $ \cite{borgesecosta2020}. Once computed the impedance matrix on the crystal axes, it can be computed also for a generic angle of incidence $(\varphi)$  as follows:

\begin{equation} \label{eq_XaxisRot}
\underline{\underline{Z}}(\varphi)  = \underline{\underline{R}}^T \underline{\underline{Z}}^{{\chi}} \underline{\underline{R}}
\end{equation}

where $ \underline{\underline{Z}}^{{\chi}} $ is the approximate impedance calculated on the crystal axes $(\theta = 0^{\circ},\varphi = \varphi^{\chi}) $ and $\underline{\underline{R}}$ is the rotation matrix:

\begin{equation} \label{eq:Rot1}
 \underline{\underline{R}} = \begin{bmatrix}
                    \cos(-\varphi^{rot}) & -\sin(-\varphi^{rot}) \\[2pt]
                    \sin(-\varphi^{rot}) & \cos(-\varphi^{rot})  \\
                  				\end{bmatrix} \\
\end{equation}

where: 

\begin{equation} \label{eq:m_rot}
 \varphi^{rot} = \varphi - \varphi^{\chi} 
\end{equation}

Once the impedance matrix of each element is computed, the reflection and transmission coefficients of a multilayer structure comprising dielectric layers and generically rotated metasurfaces can be computed according to transfer matrix (ABCD) approach \cite{pfeiffer2014bianisotropic}. Since the metasurface is anisotropic, the problem cannot be simply decomposed along $x$ and $y$ axes since TE and TM modes are coupled in general \cite{asadchy2015broadband,borgesecosta2020,maci2005pole}. When a generic azimuth angle of incidence is considered, it is more convenient to express the reflection and transmission coefficients in terms of TE and TM modes. The electric field for TM polarization is aligned with the plane of incidence $\varphi$, that is \textit{x}-axis if $\varphi=0^\circ$.  The electric field for TE polarization is aligned with the normal to the plane of incidence, that is \textit{y}-axis if $\varphi=0^\circ$.  For this reason, the metasurface impedance must be treated as a matrix.
The full scattering matrix, both for both TE and TM polarizations, can be derived as:
\begin{widetext}              				
\begin{equation}
\begin{bmatrix}
                    \underline{\underline S} _{11}^{TE/TM} &\underline{\underline S} _{21}^{TE/TM} \\[5pt]
                    \underline{\underline S} _{12}^{TE/TM} & \underline{\underline S} _{22}^{TE/TM}  
                  				\end{bmatrix}= {\begin{bmatrix}
                     - \underline{\underline I}  &  \dfrac{B \underline{\underline{n}}}{\zeta _0^{TE/TM}}  + \underline{\underline{A}}  \\[5pt]
                    \dfrac{\underline{\underline{n}}}{\zeta _0^{TE/TM}}   & \dfrac{D \underline{\underline{n}}}{\zeta _0^{TE/TM}}  + \underline{\underline{C}}   
                  				\end{bmatrix}}^{-1}                   				\begin{bmatrix}
                      \underline{\underline I}  &  \dfrac{\underline{\underline{B}} \underline{\underline{n}}}{\zeta _0^{TE/TM}}  - \underline{\underline{A}}  \\[5pt]
                    \dfrac{\underline{\underline{n}}}{\zeta _0^{TE/TM}}   & \dfrac{\underline{\underline{D}} \underline{\underline{n}}}{\zeta _0^{TE/TM}}  - \underline{\underline{C}}   
\end{bmatrix}
\end{equation}
\end{widetext}
               				
where $ \underline{\underline{I}}=\begin{bmatrix}1&0\\0&1 \end{bmatrix}$ is the identity matrix and $ \underline{\underline{n}}=\begin{bmatrix}0&-1\\1&0 \end{bmatrix}$ is the $90^{\circ}$ rotation matrix. The terms of the ABCD matrix of the cascade system is computed as follow:

\begin{equation}\label{eq:Mdiel}
\begin{bmatrix}
 \underline{\underline {\text{A}}} & \underline{\underline B}  \hfill \\
  \underline{\underline C} & \underline{\underline D}
\end{bmatrix} =  \underline{\underline{M}}_{1} \underline{\underline{M}}_{2} \underline{\underline{M}}_{3}...\underline{\underline{M}}_{i}
\end{equation} 				

where $\underline{\underline M}_i$ represents the ABCD matrix of the $i^{th}$ layer. The $ABCD$ form for the metasurface (MTS) is:

\begin{equation}
 \underline{\underline{M}}^{MTS}_i=
\begin{bmatrix}
 \underline{\underline {\text{I}}} & \underline{\underline 0}  \hfill \\
  \underline{\underline {\text{n}}} \, {\underline{\underline {\text{Y}}}_i}&\underline{\underline I}
\end{bmatrix}  
\end{equation}

where $\underline{\underline Y}_i$ is the admittance matrix ($\underline{\underline Y}_i={\underline{\underline Z}}_i^{-1}$ ), whereas the dielectric (diel) one reads:

\begin{equation}\label{eq:ABCD_diel}
\resizebox{.88\hsize}{!}{ $ \underline{\underline{M}}^{diel}_i = \begin{bmatrix}
\cos \left( {{k_{zi}}{d_i}} \right)\underline{\underline {\text{I}}} & -j \sin \left( {{k_{zi}}{d_i}} \right)\zeta _i^{TE/TM} \underline{\underline n}  \hfill \\[5pt]
  -j \dfrac{{\sin \left( {{k_{zi}}{d_i}} \right)}}{{\zeta _i^{TE/TM}}}\underline{\underline {\text{n}}} & \cos \left( {{k_{zi}}{d_i}} \right) \underline{\underline I}
                  				\end{bmatrix}$}
\end{equation}

$\zeta _0^{TE/TM}$ and $\zeta _i^{TE/TM}$ represent the impedances of the equivalent transmission line for free space and for the $i^{th}$ dielectric medium, respectively. The impedances for TE and TM incidence read:

\begin{equation}\label{eq_imepdance_TE_TM}
\zeta_i^{TE} = \frac{\omega\mu_0\mu_i}{k_{zi}}, \;\;\; \zeta_i^{TM} = \frac{k_{zi}}{\omega\varepsilon_0\varepsilon_i}
\end{equation}

where $k_{zi}$ represents the propagation constant along the normal direction inside the $i^{th}$ dielectric medium ${k_{zi}}=\sqrt{{k_0\varepsilon_i\mu_i}^2-{k_t}^2}$, with $k_t=k_0sin(\theta)$. $\varepsilon_0$ and $\mu_0$ represent the dielectric permittivity and the magnetic permeability of free space whereas
$\varepsilon_i$ and $\mu_i$ represent the relative dielectric permittivity and the relative magnetic permeability of the $i^{th}$ dielectric medium.  

The ABCD formulation can be adopted also for computing the oblique incidence behaviour of the polarization converter. However, the model does not consider angular variation (spatial dispersion) of the metasurface impedance. On the contrary, the spatially dispersive effects of the spacers are taken into account. An important rule of thumb regarding the application of the transmission line model is that the distance between the metasurfaces needs to be sufficiently large \cite{costa_efficient_2012}. 
  
\section{Numerical results} 

\begin{figure}[h!]
\centering
  \subfloat[]{%
       \includegraphics[width=0.45\linewidth]{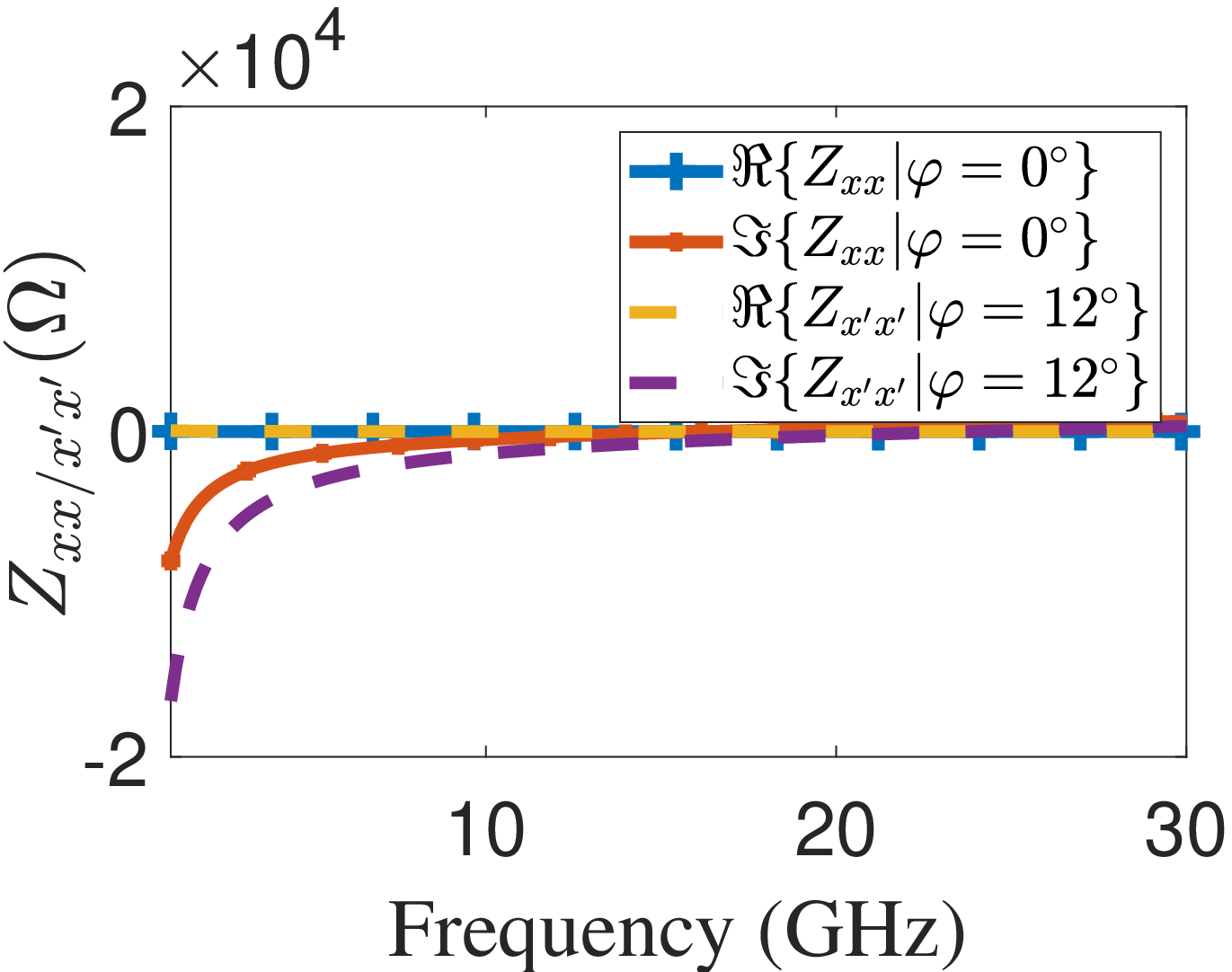}}
\quad
  \subfloat[]{%
        \includegraphics[width=0.45\linewidth]{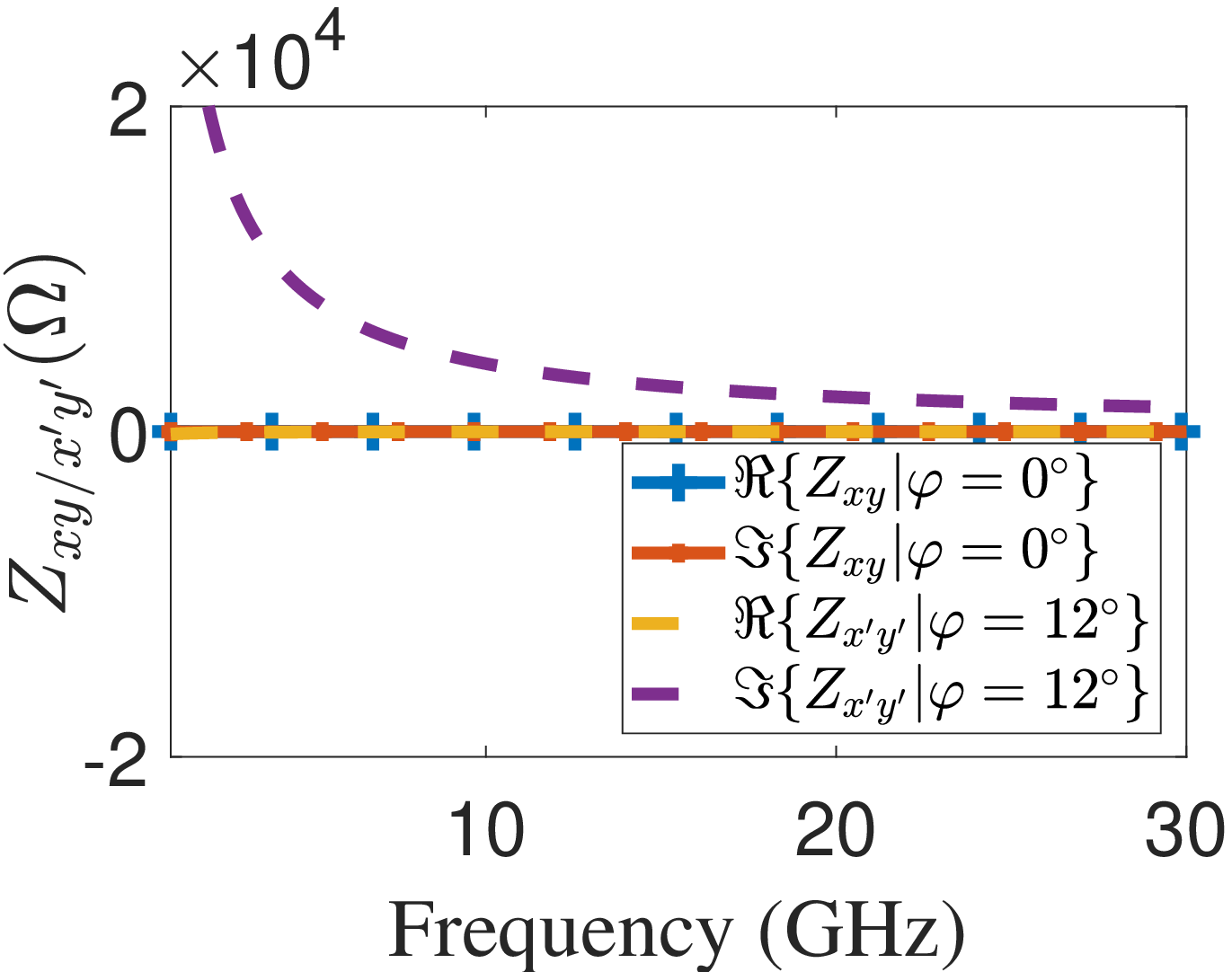}}
\quad
  \subfloat[]{%
        \includegraphics[width=0.45\linewidth]{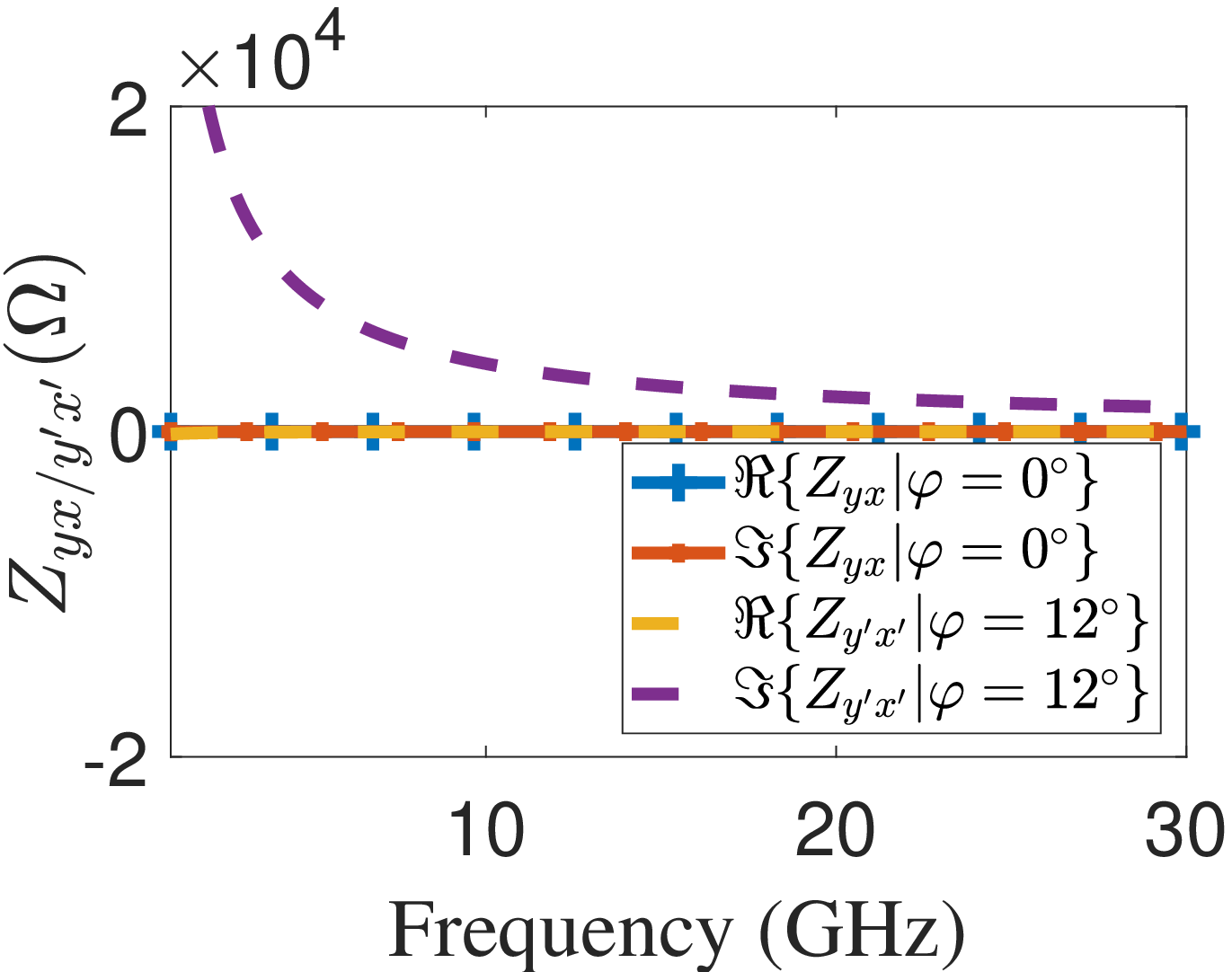}}
\quad
  \subfloat[]{%
        \includegraphics[width=0.45\linewidth]{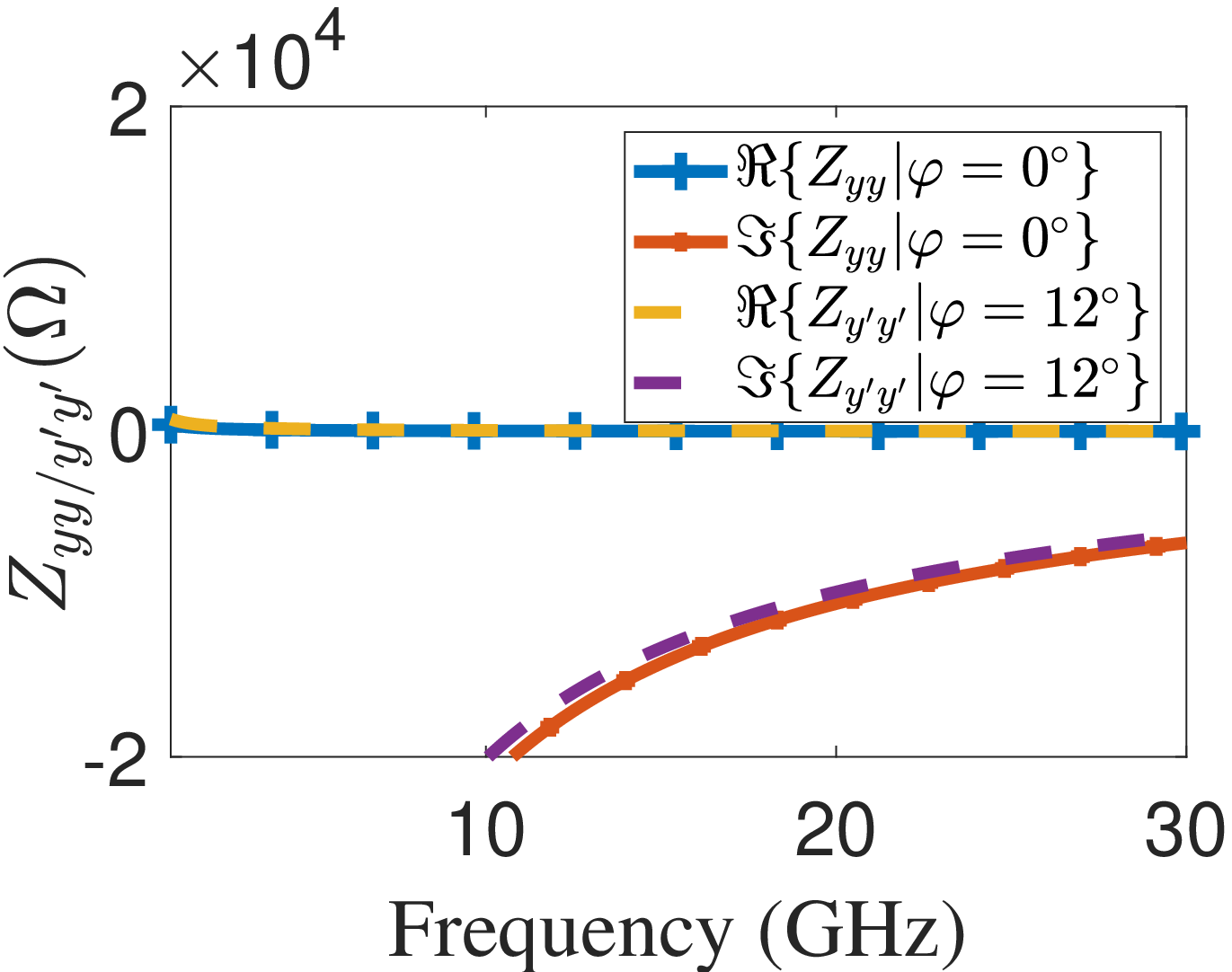}}
      \caption{$ \Re{\lbrace\underline{\underline{Z}}}\rbrace $ and $ \Im{\lbrace\underline{\underline{Z}}}\rbrace $ as a function of the frequency for the first and second metasurface layers. The matrix $\underline{\underline{Z}}$ is represented on Cartesian axes $(x,y)$ and $(x^\prime,y^\prime)$ where the latter is the rotated reference system. The elevation angle of the impinging electric field is $ (\theta = 0^{\circ} ) $. The metasurface is interrogated with two different azimuth angles $ (\varphi = 0^{\circ}, 12^\circ)$; (a) $ Z_{xx/x^{\prime}x^{\prime}} $, (b) $ Z_{xy/x^{\prime}y^{\prime}} $, (c) $ Z_{yx/y^{\prime}x^{\prime}} $,(d) $ Z_{yy/y^{\prime}y^{\prime}} $. The metasurface unit cell is the dipole shown in Fig.~\ref{fig_UnitCells}(b) with a periodicity of $D=10$ mm a length of 8.75 mm and width of 1.25 mm.}
  \label{fig_ZFSS_dipoles_1_2} 
\end{figure}

\begin{figure}[h!]
\centering
  \subfloat[]{%
       \includegraphics[width=0.45\linewidth]{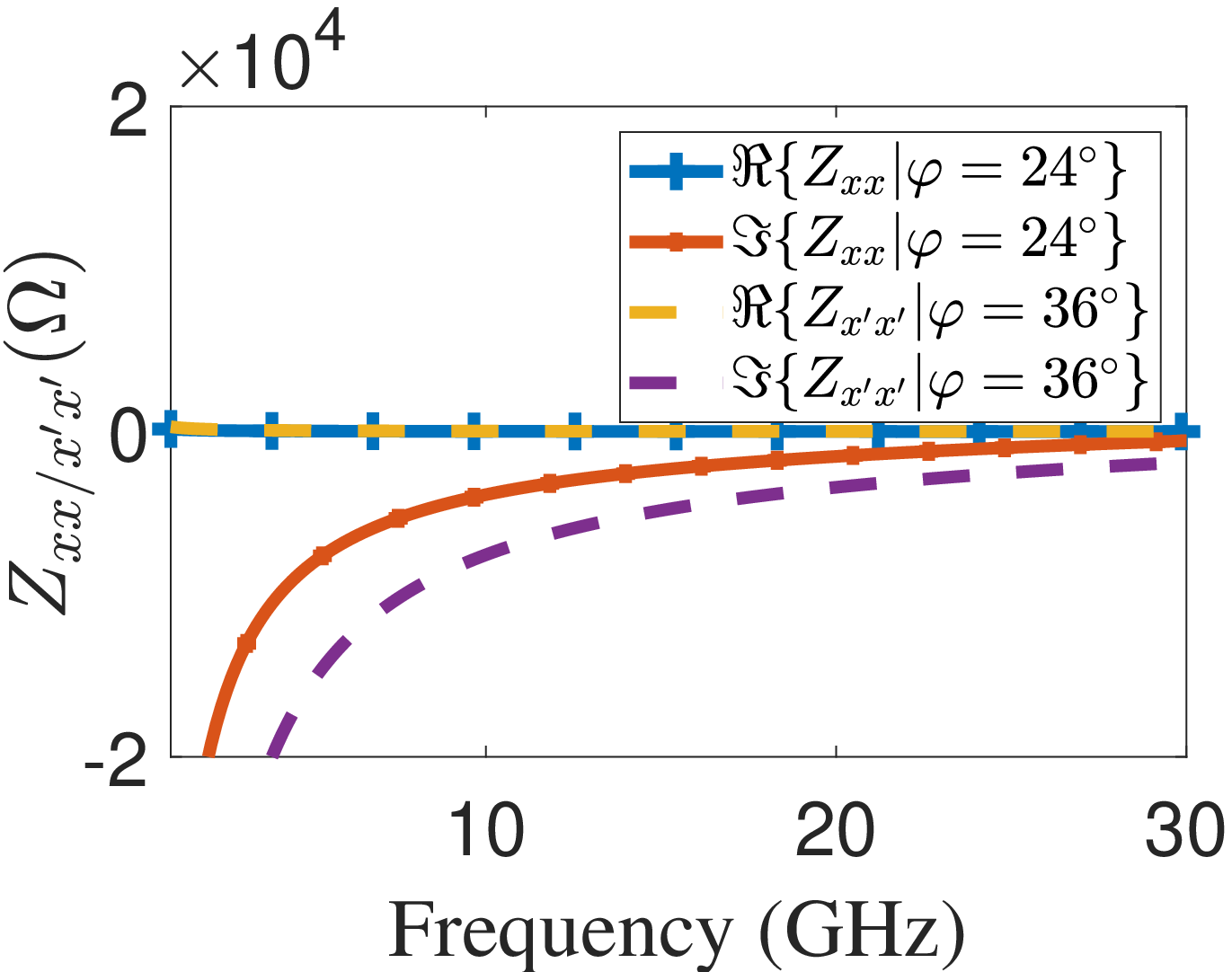}}
\quad
  \subfloat[]{%
        \includegraphics[width=0.45\linewidth]{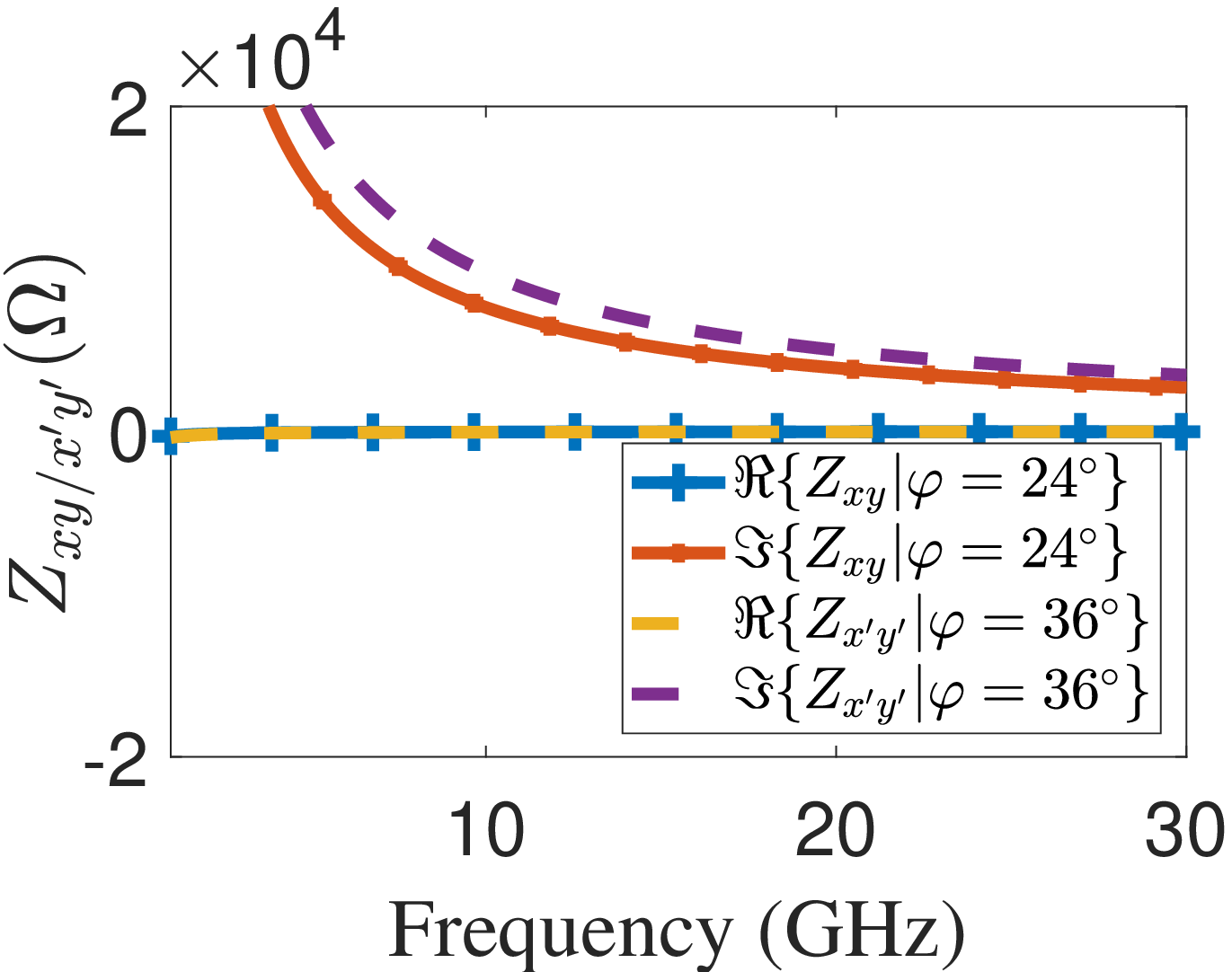}}
\quad
  \subfloat[]{%
        \includegraphics[width=0.45\linewidth]{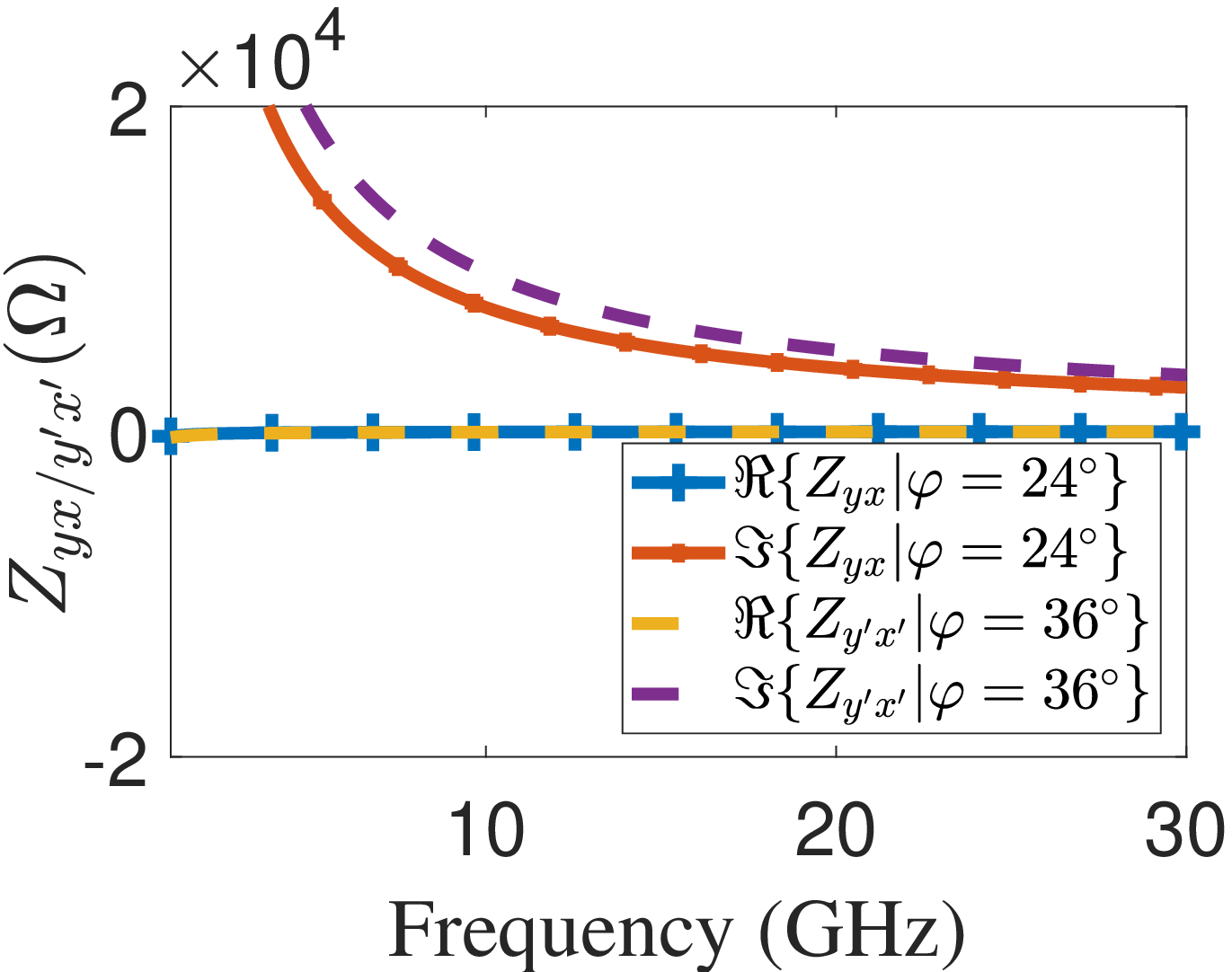}}
\quad
  \subfloat[]{%
        \includegraphics[width=0.45\linewidth]{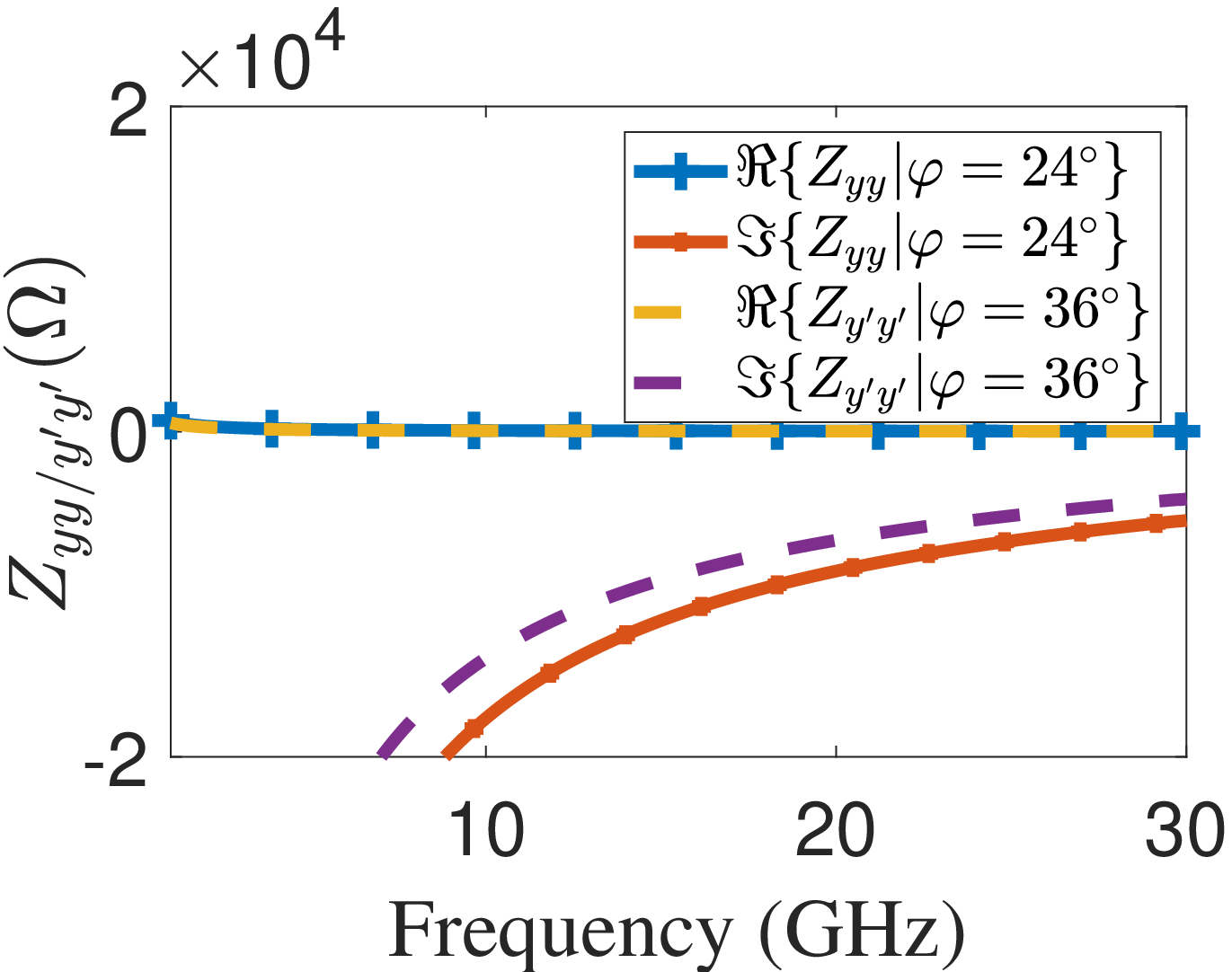}}
      \caption{$ \Re{\lbrace\underline{\underline{Z}}}\rbrace $ and $ \Im{\lbrace\underline{\underline{Z}}}\rbrace $ as a function of the frequency for the third and forth metasurface layers. The matrix $\underline{\underline{Z}}$ is represented on Cartesian axes $(x,y)$ and $(x^\prime,y^\prime)$ where the latter is the rotated reference system. The elevation angle of the impinging electric field is $ (\theta = 0^{\circ} ) $. The metasurface is interrogated with two different azimuth angles $ (\varphi = 24^{\circ}, 36^\circ)$; (a) $ Z_{xx/x^{\prime}x^{\prime}} $, (b) $ Z_{xy/x^{\prime}y^{\prime}} $, (c) $ Z_{yx/y^{\prime}x^{\prime}} $,(d) $ Z_{yy/y^{\prime}y^{\prime}} $. The metasurface unit cell is the dipole shown in Fig.~\ref{fig_UnitCells}(b) with a periodicity of $D=10$  mm a length of 8.75 mm and width of 1.25 mm.}
  \label{fig_ZFSS_dipoles_3_4} 
\end{figure}

\begin{figure} 
\centering
  \subfloat[]{%
\includegraphics[width=0.45\linewidth]{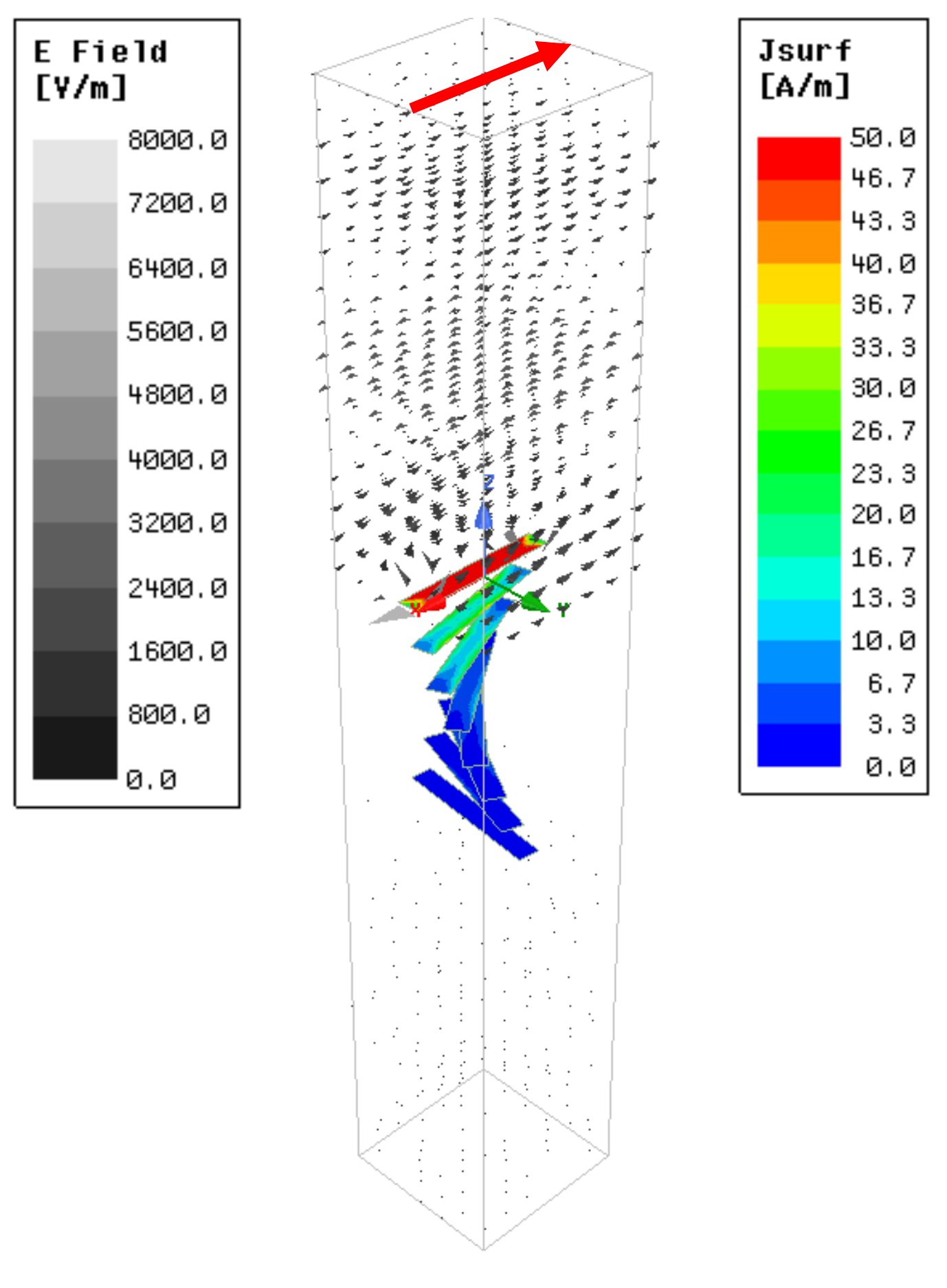}}
\quad
  \subfloat[]{%
        \includegraphics[width=0.45\linewidth]{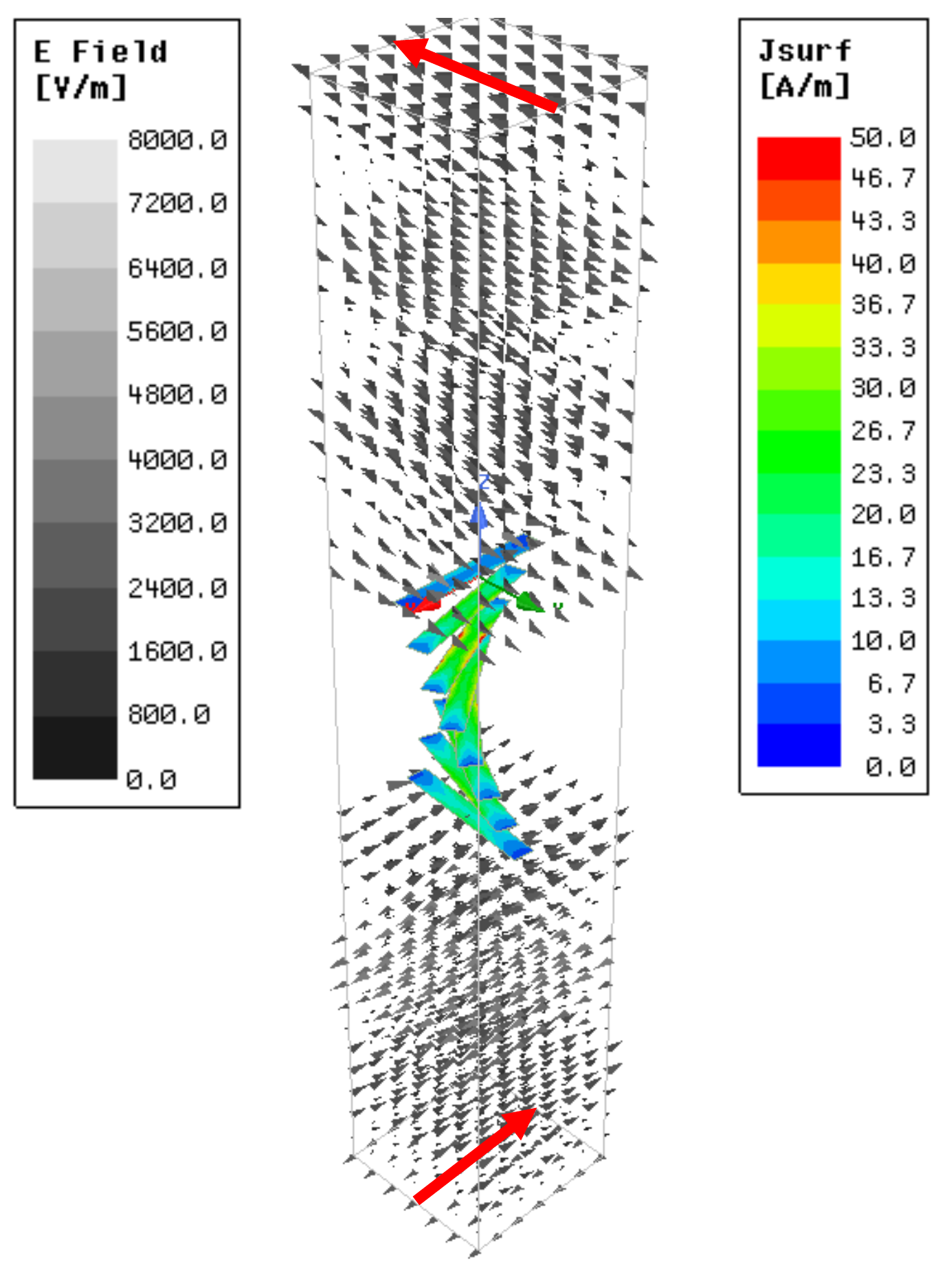}}
       \caption{Surface currents and electric fields in the unit cell of the multilayer polarization converter (a) \textit{TM} excitation , (b) \textit{TE} excitation}
  \label{fig_currents_fields} 
\end{figure}

\begin{figure} 
\centering
  \subfloat[]{%
       \includegraphics[width=0.45\linewidth]{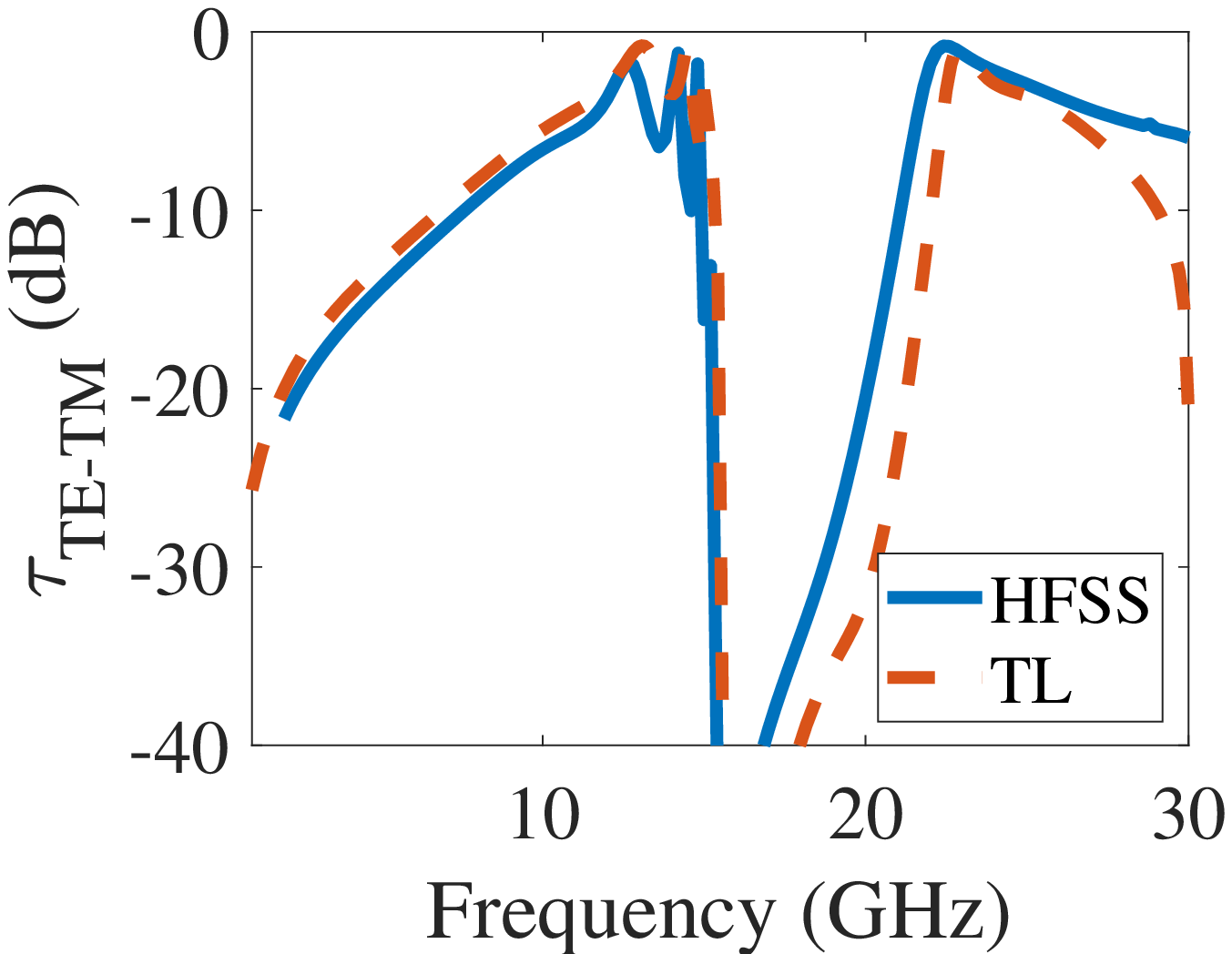}}
\quad
  \subfloat[]{%
        \includegraphics[width=0.45\linewidth]{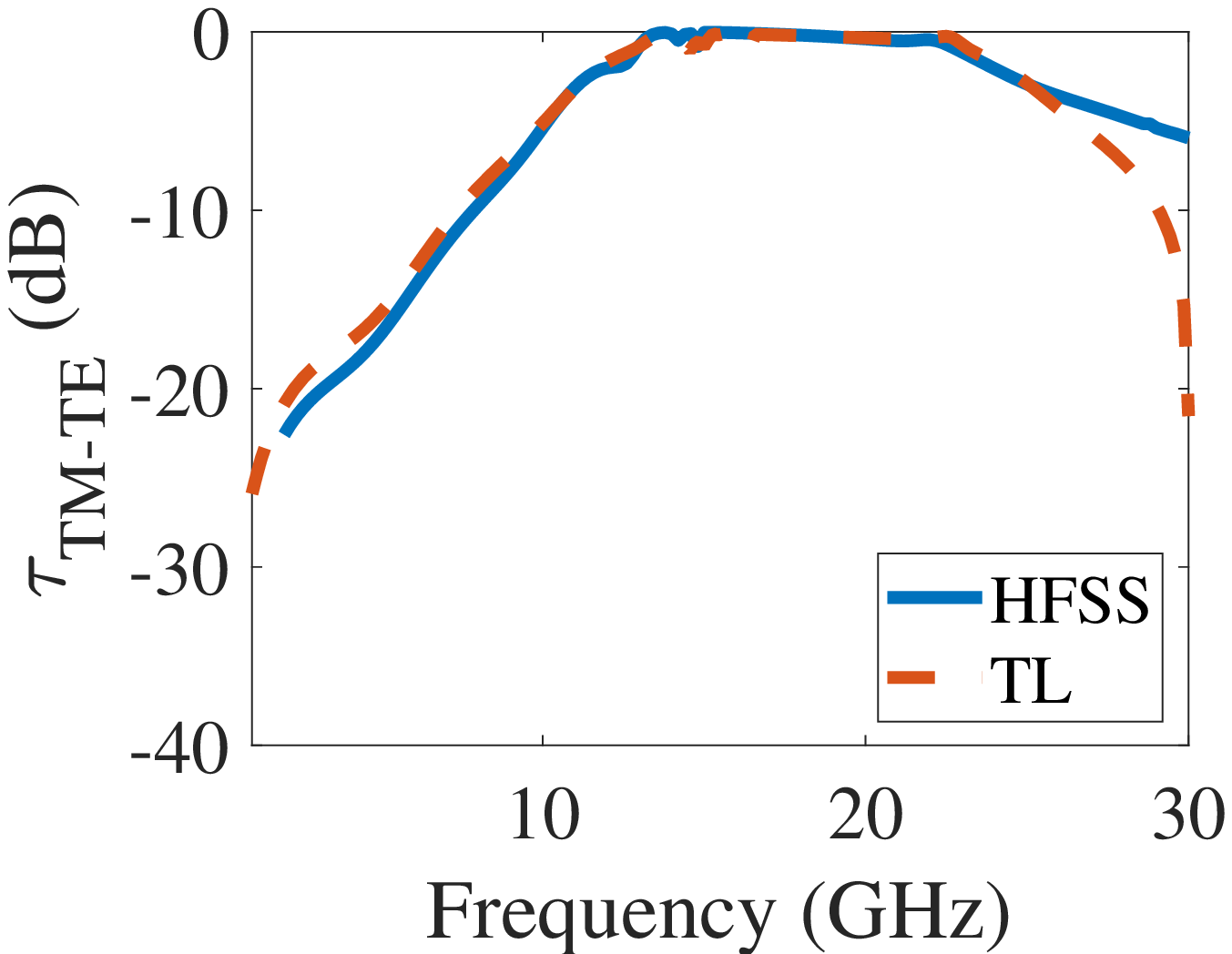}}
       \caption{Performance of the dipole based multilayer polarization converter. The optimal configuration comprises 8 layers spaced by 2 mm air spacers. (a) $\tau_{TETM}$, (b) $\tau_{TMTE}$.}
  \label{fig_tx_crosspolar} 
\end{figure}

The polarization converter analysis has been carried out by considering both capacitive and inductive metasurfaces. Among the considered elements geometries shown in Fig.~\ref{fig_UnitCells}, the best compromise between thickness and operating bandwidth has been obtained by using 8 layers of progressively rotated dipoles.  The dipoles are gradually rotated by an angle of $12^\circ$ and are separated by $2$ mm of air. The dipole element (Fig.~\ref{fig_UnitCells} (a)) is characterized by a periodicity $D = 10$ mm towards both planar directions. The length of the dipole is 8.75 mm and its width is 1.25  mm. The performance of the structure analyzed by using the analytical ABCD formulation have been verified by using a full-wave electromagnetic simulation with Ansys HFSS as shown in Fig.~\ref{fig_currents_fields}. The impedance behaviour of the dipole metasurface for the first two layers is reported in Fig.~\ref{fig_ZFSS_dipoles_1_2}. The impedance behaviour of the dipole metasurface for the third and forth layer are reported in Fig.~\ref{fig_ZFSS_dipoles_3_4}. It can be observed that the impedance of the first layer is diagonal since the element is aligned with its crystal axis. As the rotation is applied, the resonance frequency of the $x^{\prime}x^{\prime}$-term of the impedance moves progressively towards higher frequency showing a capacitive behaviour inside the polarization conversion frequency band. At the same time, the off-diagonal terms of the impedance matrix start having a non-negligible positive imaginary part. For the analysed dipole topology, the impedance of the rotated elements inside the square lattice is comparable to the impedance of the rotated screen thus justifying the setup employed for the EM simulation. The Fig.~\ref{fig_tx_crosspolar} reports the cross-polar reflection coefficient for the optimal polarization converter. In Fig.~\ref{fig_tx_crosspolar}(a), the transmission conversion from TE polarization to TM polarization is reported whereas Fig.~\ref{fig_tx_crosspolar}(b) reports the transmission conversion from TM polarization to TE polarization.

\begin{figure}
\centering
  \subfloat[]{%
 		\includegraphics[width=0.45\linewidth]{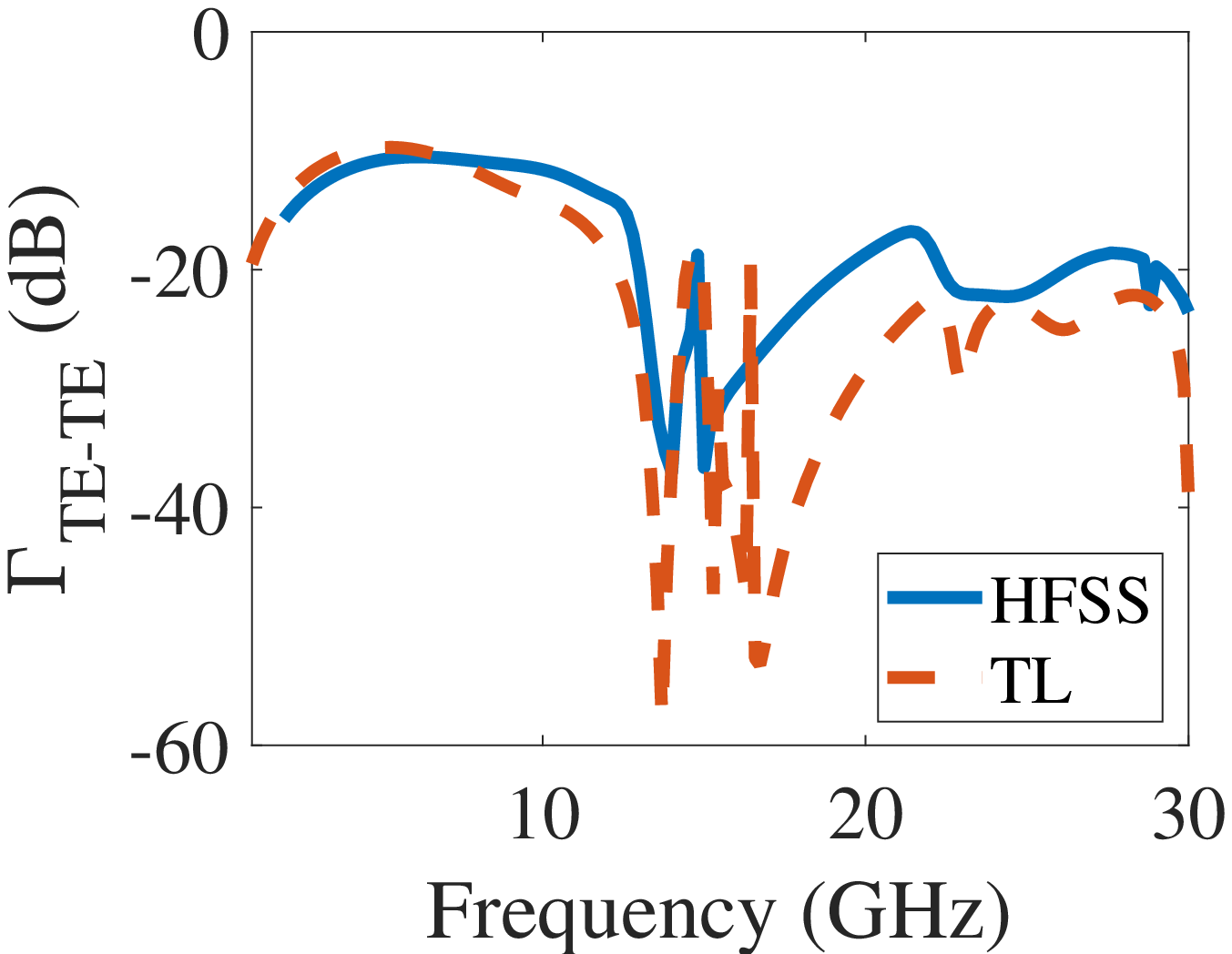}}
\quad
  \subfloat[]{%
        \includegraphics[width=0.45\linewidth]{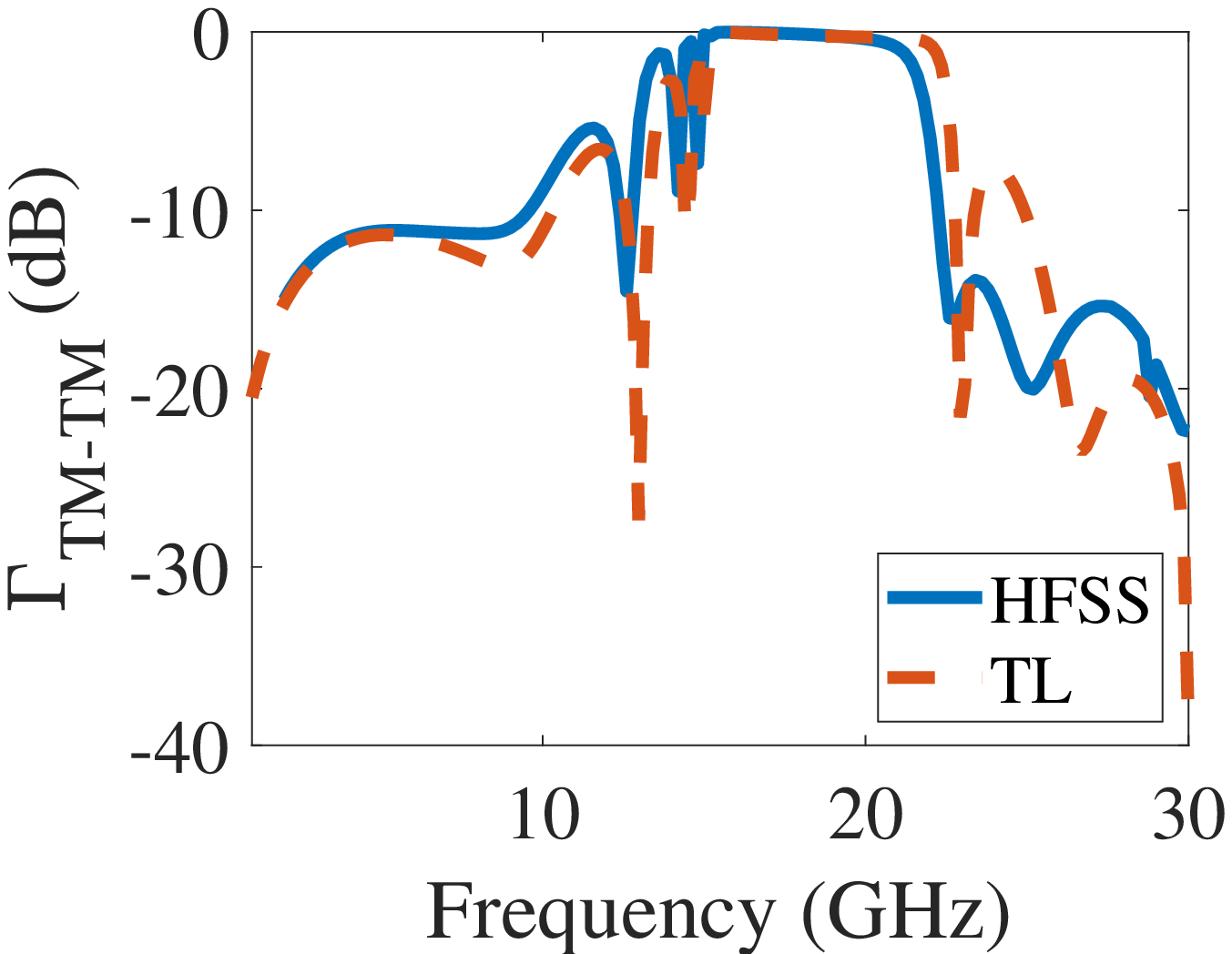}}
\quad
  \subfloat[]{%
       \includegraphics[width=0.45\linewidth]{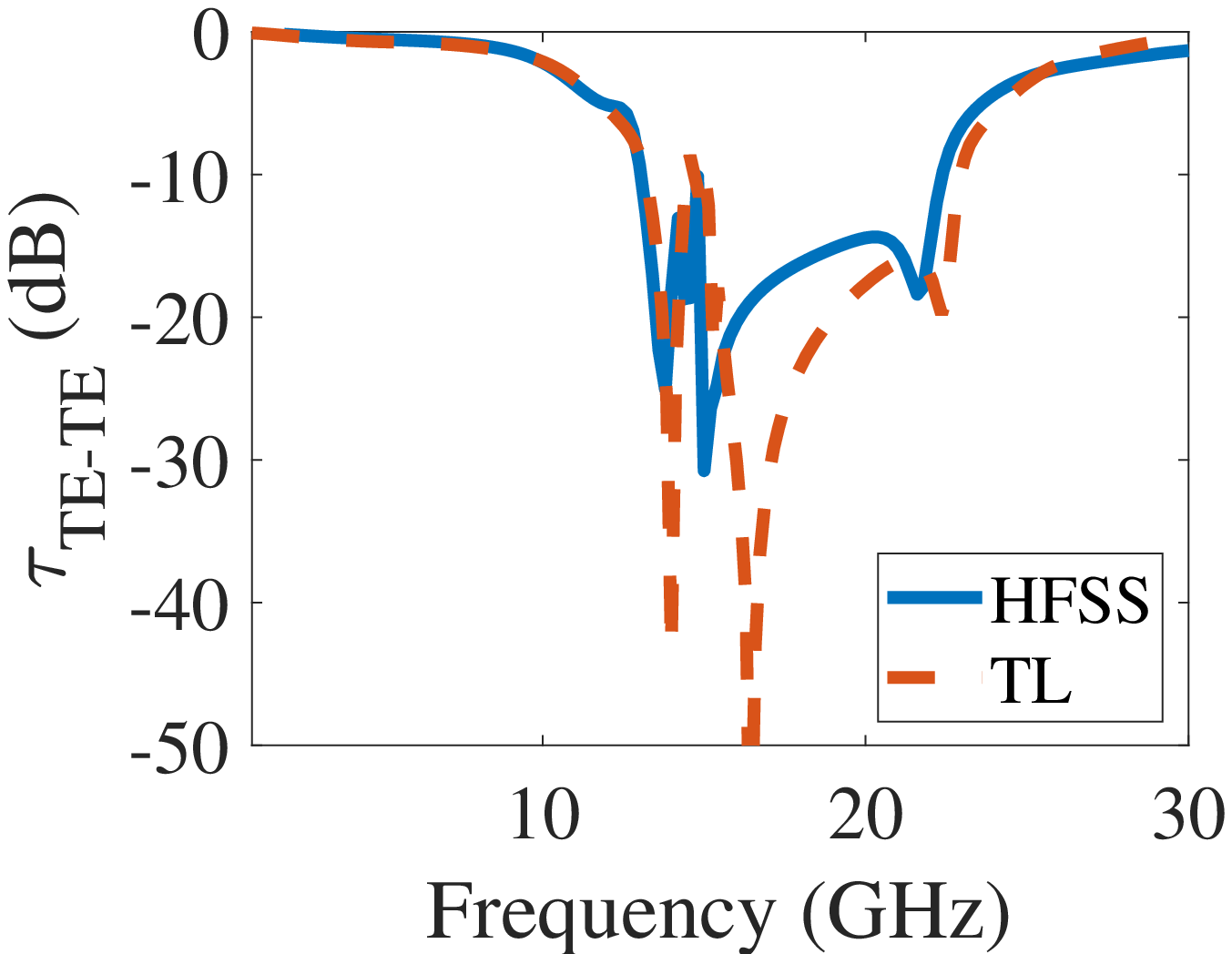}}
\quad
 \subfloat[]{%
        \includegraphics[width=0.45\linewidth]{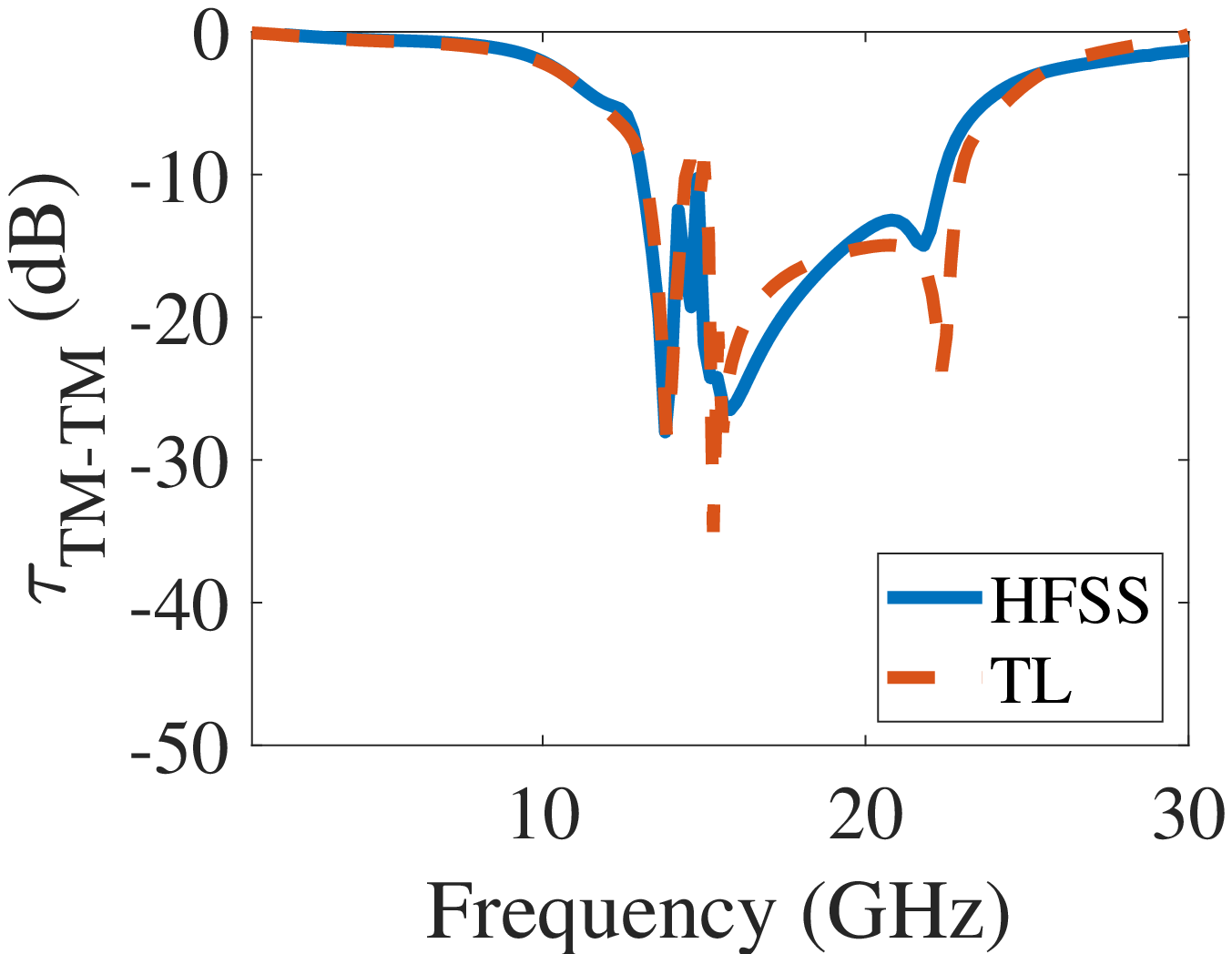}}
      \caption{Performance of the dipole based multilayer polarization converter. The optimal configuration comprises 8 layers spaced by 2 mm air spacers. (a) $\Gamma_{TETE}$, (b) $\Gamma_{TMTM}$, (c) $\tau_{TETE}$,(d) $\tau_{TMTM}$. }
  \label{fig_rx_tx_copolar} 
\end{figure}

\begin{figure} 
\centering
  \subfloat[]{%
       \includegraphics[width=0.45\linewidth]{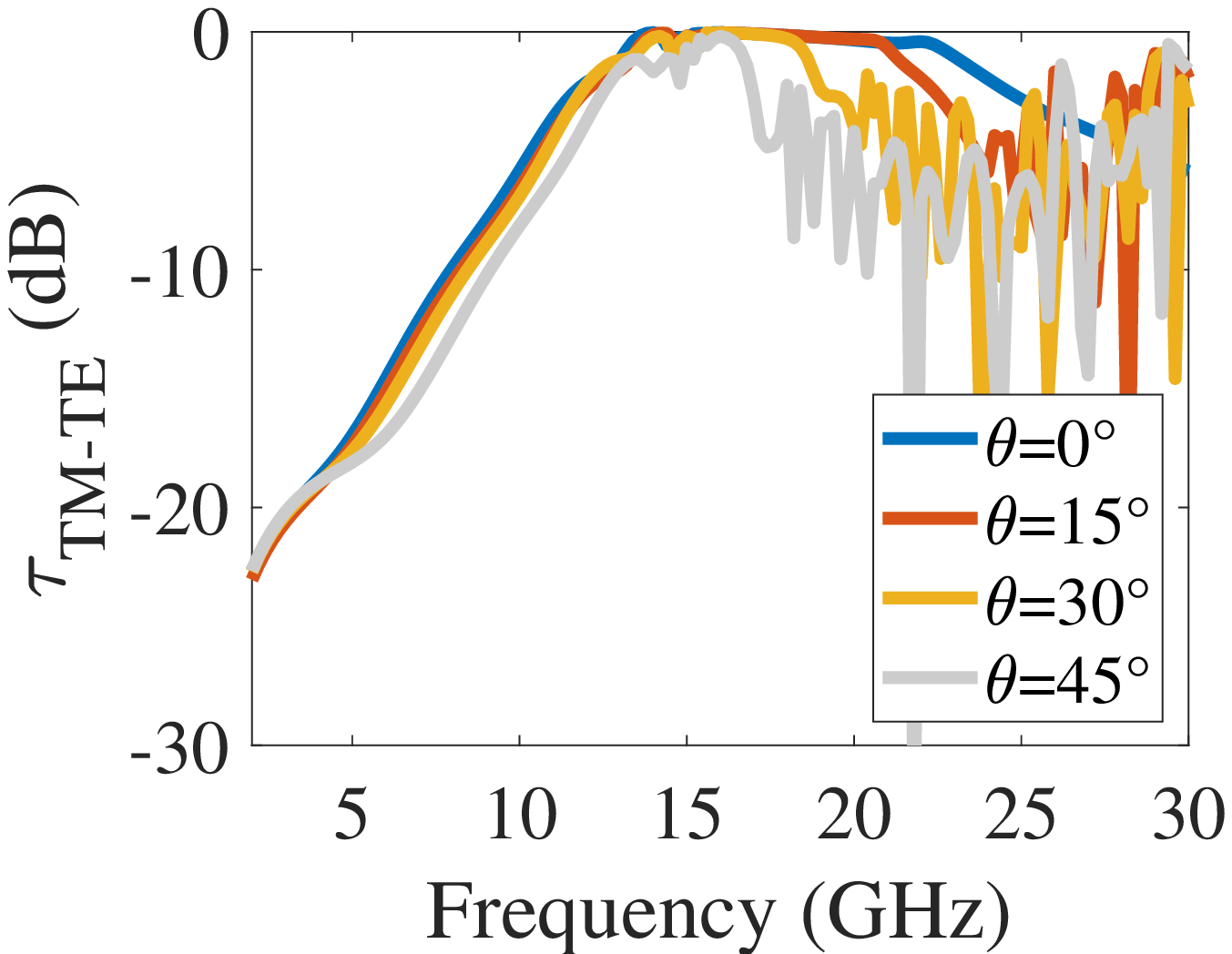}}
\quad
  \subfloat[]{%
        \includegraphics[width=0.45\linewidth]{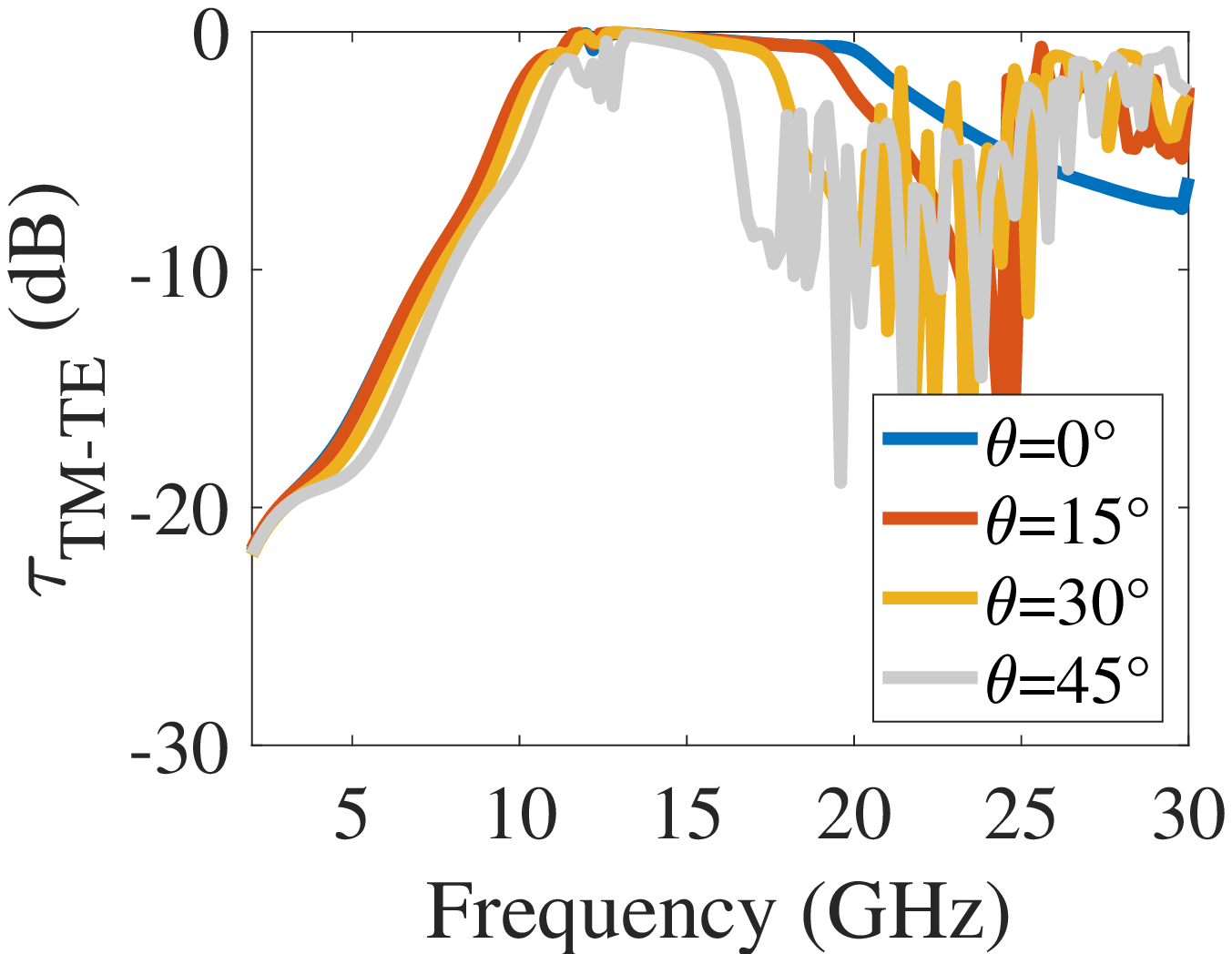}}
       \caption{Performance of the polarization converters as a function of the elevation angle of incidence $\theta$: (a) sol. [610], (b) sol. [1110]. The simulations have been performed with Ansys HFSS.}
  \label{fig_oblique_incidence} 
\end{figure}

As evident, the behaviour of the polarization converter is asymmetric as only the TE polarized fields are converted into the other polarization while the TM polarized ones are reflected. The impedances of the metasurface layers in Fig.~\ref{fig_ZFSS_dipoles_1_2} and in Fig.~\ref{fig_ZFSS_dipoles_3_4}, show that the $y$-component of the metasurface impedance, which interferes with the TE fields, is largely capacitive and thus not reflecting while the $x$-component, which interferes with TM fields, is resonant and thus reflective around 18 GHz. For this reason, the TE fields are transmitted and gradually rotated by the polarization converter whereas the TM fields are strongly reflected by the first layer or the other elements of the metallic surface. To confirm the asymmetric behaviour of the polarization converter, both the surface currents and the electric fields on the unit cell of the multilayer polarization converter at 18 GHz are shown in  Fig.~\ref{fig_currents_fields}. The field distributions have been obtained by using Ansys HFSS. It is evident that when the multilayer structure is excited with TE polarization, the fields go through the multilayer structure and it is transformed into TM polarized fields. On the contrary, when the structure is excited with TM polarization, the fields are completely reflected by the first layer of the multilayer structure.
The co-polar transmission and reflection coefficients are reported In Fig.~\ref{fig_rx_tx_copolar}. The reflection coefficient is low for the co-polar component \textit{TE-TE} while a large reflection is achieved for other co-polar component \textit{TM-TM}. The co-polar transmission coefficients are  low for both \textit{TE-TE} and \textit{TM-TM} components. The performance of the polarization converter at oblique incidence is shown in Fig.~\ref{fig_oblique_incidence}. The solution with the dipole unit cell ([610]) is characterized by a cell periodicity of 10 mm and it operates polarization conversion up-to 25 GHz at normal incidence. Therefore, as the incidence angle increases, the high order harmonics start propagating inside the operative band of the polarization converter leading to a bandwidth reduction. We report in Fig.~\ref{fig_oblique_incidence} also the performance of the solution with the loaded dipole ([1110]) shown in Fig.~\ref{fig_UnitCells}(c) which is characterized by a similar percentage bandwidth at normal incidence with the same periodicity. However the solution [1110] operates polarization conversion at lower frequencies thus leading to a more compact configuration in terms of periodicity over wavelength. For this reason, the performance of the solution [1110] are better than the solution [610] at oblique incidence. The percentage bandwidth, $BW_\alpha (\%)$ ($\alpha=-1$), of the two configurations as a function of the incidence angle are summarized in Table ~\ref{tab_oblique_inc_BW}. However, it is worth to point out that the performance of the polarization converters presented in the paper are not optimized for oblique incidence. 

{\setlength{\extrarowheight}{3pt}%
\begin{table}[ht]
\caption{Percentage bandwidth,$BW_\alpha (\%)$, of the solutions [610] and [1110] as a function of the incidence angle $\theta$.}
\centering
\begin{tabularx}{\columnwidth}{|X|X|X|X|}
\hline
\textbf{\; Sol. idx} & \hspace{16pt}$\boldsymbol{\theta=0^\circ$} & \hspace{16pt}$\boldsymbol{\theta=15^\circ$} & \hspace{16pt}$\boldsymbol{\theta=30^\circ$}\\[3pt]
\hline
\hspace{16pt}[610] &  \hspace{16pt}$53.3 \% $ &   \hspace{16pt}$43.7 \% $ &   \hspace{16pt}$21.73 \% $ \\[3pt]
\hline
\hspace{16pt}[1110] & \hspace{16pt}$55.7 \% $ &  \hspace{16pt}$54.3 \% $ & \hspace{16pt}$44.0 \% $ \\[3pt]
\hline
\end{tabularx}
\label{tab_oblique_inc_BW}
\end{table}

\section{Measured results} 

A prototype of the 8-layers polarization converter comprising dipole resonators has been fabricated and measured. The 8 layers have been manufactured on a thin Kapton (35 $\mu$m) substrate with a standard photo-lithographic technology. The metasurfaces have been assembled by separating the layers with 2 mm spacer realized with Rohacell. The metasurfaces and the assembled prototype are shown in Fig.~\ref{fig_measures}(a). The planar dimensions of the sample are 13 cm $\times$ 13 cm. The total thickness is equal to 14 mm. Two dual-polarized wideband horn antennas (Flann DP280) are used to interrogate the sample and to receive the cross-polar transmitted field on the other side of the sample. The gain of the reference antennas is approximately 8 dBi at 2 GHz and it increases up to 13.3 dBi at 8 GHz. An Agilent E5071C vector network analyser is employed to measure the scattering parameters. The two wideband horn antennas have been placed at  distance of 1 m. The measuring setup is shown in Fig.~\ref{fig_measures}(a). Initially, the transmitted field level is measured without the polarization converter with the antennas fed with the same polarization. The alignment has been performed by looking at the maximum received power with co-polar component. This value represents the reference field, that is, the maximum receivable electric field. Once performed the alignment of the antennas, an electromagnetic transparent polystyrene sample holder has been interposed between the antennas. After this calibration measurement, the polarization of the receiving antenna is switched from vertical to horizontal and much lower level of field is clearly received. When the polarization converter is placed between the antennas, the received field level is restored to the reference values within the polarization converting band. In particular, the measured value (in decibel) is subtracted from the reference value to obtain the normalized curve. The measured polarization conversion is shown in Fig.~\ref{fig_measures}(b) where the simulated cross-polar transmission coefficient obtained from full-wave simulations is also reported. Even if the sample size covers only the -3dB beamwidth of the horn antenna, the agreement between measurements and simulations is good.

\begin{figure} 
\centering
  \subfloat[]{%
       \includegraphics[width=0.45\linewidth]{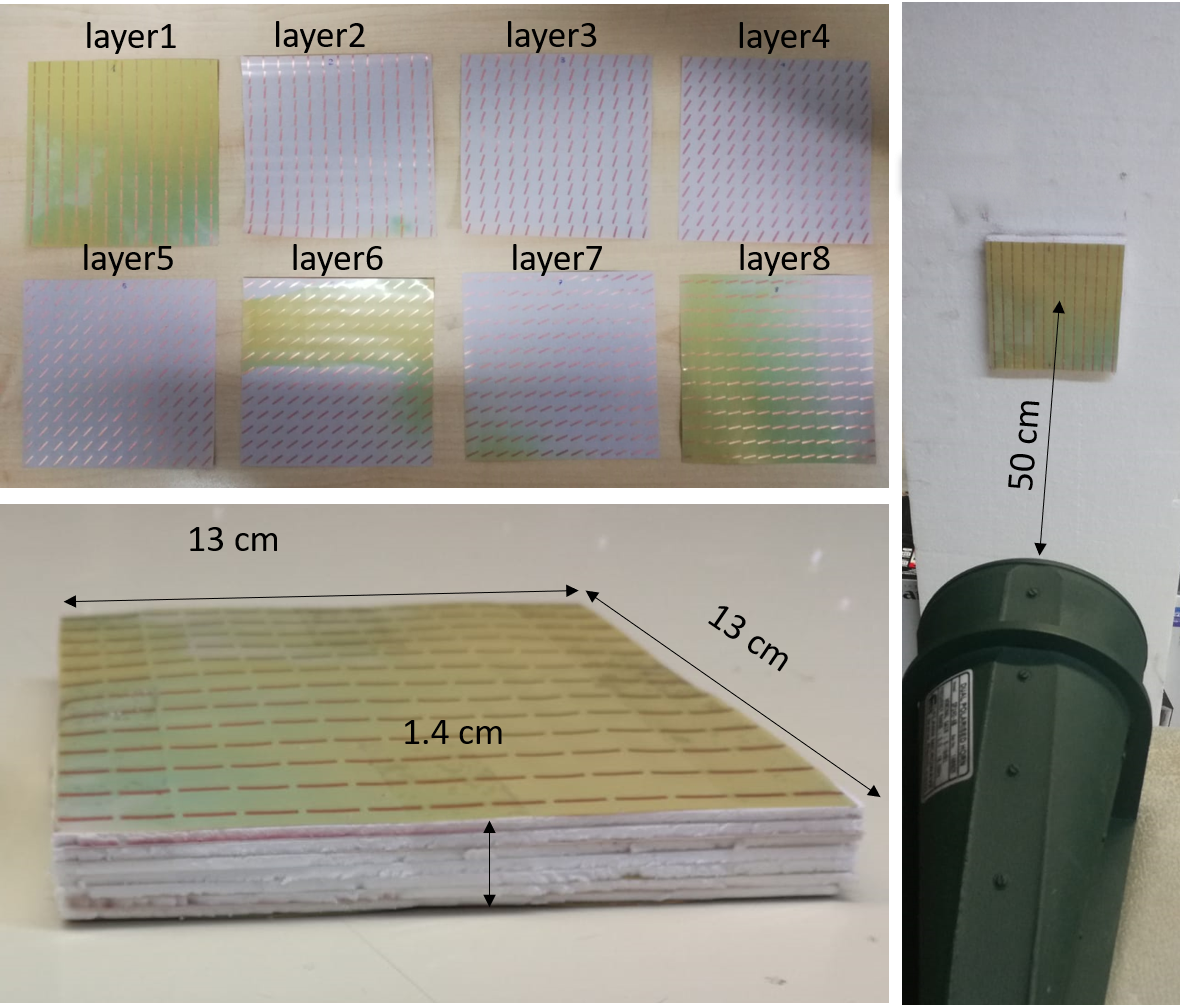}}
 \quad
  \subfloat[]{%
        \includegraphics[width=0.45\linewidth]{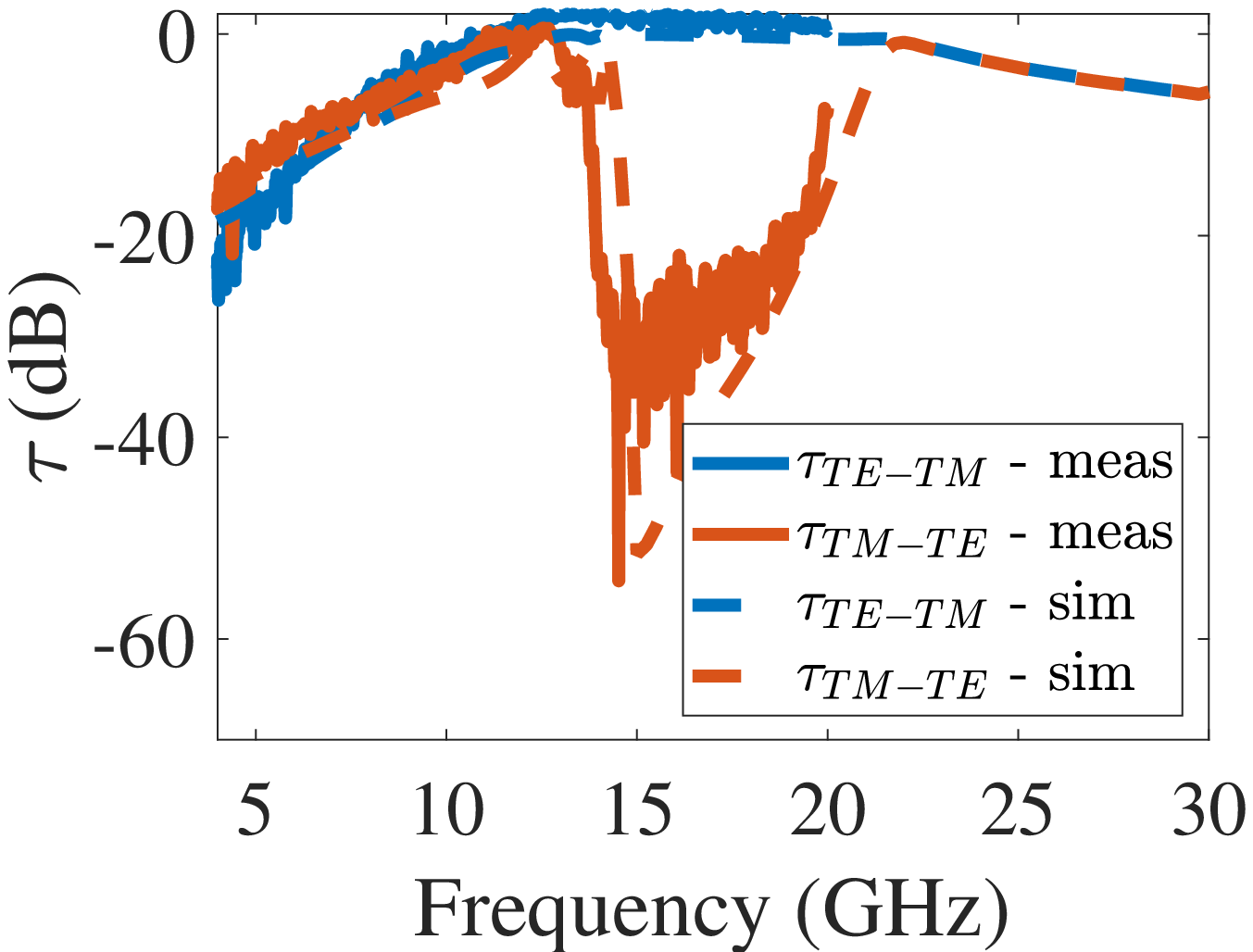}}
       \caption{(a) Pictures of the fabricated surfaces and of assembled polarization converter prototype. (b) Measured cross-polar transmission coefficient for the wideband polarization converter comprising 8-layers of gradually rotated dipoles. The dipole length is equal to 8.75 mm and its width is 1.25 mm. The dipole rotation angle layer by layer is $\varphi=12^\circ$.}
  \label{fig_measures} 
\end{figure}

\section{Conclusion}

A systematic methodology for the design of transmission-type polarization converters has been presented. The proposed approach has been applied in the design of asymmetric polarization converters synthesized by progressively rotating an anisotropic element. The cross-polarized transmitted field is analytically computed  by employing an ABCD matrix formulation for multilayer anisotropic surfaces. The impedance matrix for each metasurface layer is derived by resorting to a spectral rotation. The results have been verified by using full-wave electromagnetic simulations. The multilayer structure has been fabricated and the performance experimentally measured showed a good agreement with respect to expected results.

\appendix

\section{Crystal axis of a metasurface}
Being the metasurface  a passive system, it is possible to demonstrate through the Spectral theorem that its transmission coefficient matrix is diagonal:
\begin{equation}
{\underline{\underline R} ^{ - 1}}\underline{\underline{\tau}} \;\underline{\underline R}  = \underline{\underline D}
\end{equation}

with $\underline{\underline{R}}$ defined as:

\begin{equation} \label{eq:Rot}
 \underline{\underline{R}} = \begin{bmatrix}
                    \cos(-\varphi^{rot}) & -\sin(-\varphi^{rot}) \\[2pt]
                    \sin(-\varphi^{rot}) & \cos(-\varphi^{rot})  \\
                  				\end{bmatrix} \\
\end{equation}

It is convenient to write the rotation matrix $ \underline{\underline{R}} $ as a function of the real parameter $ m $:
\begin{equation}\label{eq:RotMatrixparam}
\underline{\underline R}  =   \frac{1}{\sqrt{1 + m^2}}  \begin{bmatrix}
                   1 & -m \\[3pt]
                   m &  1  \\
                  				\end{bmatrix} \\ 
\end{equation}

It is worth noticing that the parametric form of $ \underline{\underline{R}} $ reported in equation \eqref{eq:RotMatrixparam} exhibits the typical properties of the rotation matrix:
\begin{equation} \label{eq:Diagonalization}
\underline{\underline R}^{T}\underline{\underline R} = \underline{\underline{I}} \,\,\,\, \text{and} \,\,\,\, det(\underline{\underline{R}})=1
\end{equation}
Using relations \eqref{eq:RotMatrixparam} and \eqref{eq:Diagonalization} the following equation is derived:
\begin{equation}
{\underline{\underline R} ^{ - 1}}\underline{\underline{\tau}} \;\underline{\underline R}  =  {\underline{\underline R} ^T}\underline{\underline{\tau}} \;\underline{\underline R} = \underline{\underline{D}}
\end{equation}
Imposing that $ \underline{\underline{D}} $ is diagonal, the following relation is obtained:
\begin{equation}\label{eq:DiagZero}
\tau_{xy}-m\,\tau_{xx}+m\, \tau_{yy}- m^2\, \tau_{xy}= 0
\end{equation}
From equation \eqref{eq:DiagZero}, the parameter $ m $ can be computed:

\begin{equation}
m = \frac{\tau _{yy} \pm \tau _{xx} + \sqrt {{{\left( {\tau _{yy} - \tau _{xx}} \right)}^2} + 4{\tau _{yx}^2}}}{2\tau _{yx}}
\end{equation}

Finally, the rotation angle can be calculated as follows: 

\begin{equation}
\varphi^{rot} = \arcsin\left( m \over{\sqrt{1+m^2}} \right) 
\end{equation}

\nocite{*}

\bibliography{FINAL_VERSION}

\end{document}